\documentclass[nofootinbib]{revtex4}
\usepackage{graphicx}% Include figure files
\usepackage{dcolumn}% Align table columns on decimal point
\usepackage{amsmath,amssymb,epsfig}
\usepackage{paralist}
\usepackage{comment}
\usepackage{graphicx}
\usepackage{wrapfig}
\usepackage{multirow}
\usepackage{color,soul}
\usepackage{suffix}
\usepackage{mathtools}
\usepackage{hyperref}

\allowdisplaybreaks

\renewcommand{\vec}[1]{\boldsymbol{\mathrm{#1}}}

\begin{document}

\title{
Gravitational lensing by an extended mass distribution}

\author{Slava G. Turyshev$^{1}$, Viktor T. Toth$^2$}

\affiliation{\vskip 3pt
$^1$Jet Propulsion Laboratory, California Institute of Technology,\\4800 Oak Grove Drive, Pasadena, CA 91109-0899, USA}

\affiliation{\vskip 3pt$^2$Ottawa, Ontario K1N 9H5, Canada}

\date{\today}

\begin{abstract}

We continue our investigation of the optical properties of the solar gravitational lens (SGL).  We treat the Sun as an extended axisymmetric body and model its gravitational field using zonal harmonics. We consider a point source that is positioned at a large but finite distance from the Sun and, using our new angular eikonal method, we established the electro-magnetic (EM) field on the image plane in the focal region behind the SGL and derive the SGL's impulse response in the form of its point-spread function (PSF). The expression that we derive describes the extended Sun in all regions of interest, including the regions of strong and weak interference and the region of geometric optics. The result is in the form of a single integral with respect to the azimuthal angle of the impact parameter, covering all lensing regimes of the SGL. The same expression can be used to describe gravitational lensing by a compact axisymmetric mass distribution, characterized by small deviations from spherical symmetry. It is valid in all lensing regimes. We also derive results that describe the intensity of light observed by an imaging telescope in the focal region. We present results of numerical simulations showing the view by a telescope that moves in the image plane toward the optical axis. We consider imaging of both point and extended sources. We show that while point sources yield a number of distinct images consistent with the caustics due to zonal
harmonics of a particular order (e.g., Einstein cross), extended sources always result in the formation of an Einstein ring. These results represent the most comprehensive wave-theoretical treatment of gravitational lensing in the weak gravitational field of a compact axisymmetric gravitating object.

\end{abstract}

%\pacs{03.30.+p, 04.25.Nx, 04.80.-y, 06.30.Gv, 95.10.Eg, 95.10.Jk, 95.55.Pe}
%\keywords{}

\maketitle

\section{Introduction}

Gravitational lensing is recognized as a unique tool to conduct many important investigations in modern astrophysics \cite{Liebes:1964,Schneider-Ehlers-Falco:1992,Schneider-etal:2006}. Since the beginning of the 21st century, it is used to study the distribution of matter in stelar structures, to probe the dark matter distribution in the universe, even to search for exoplanets orbiting distant stars \cite{Refsdal:1964,Blandford-Narayan:1992,Wambsganss:1998,Gaudi:2012}.

Due to the nonlinear nature of the equations involved, most relevant efforts were constrained to monopole lenses, where the gravitational field is taken to be that of a structureless point source \cite{Herlt-Stephani:1976,Deguchi-Watson:1986,Narayan-Bartelmann:1996}. Nevertheless, there have been attempts to model extended lenses, including quadrupole and general shear distortions of the lensing potential \cite{Kovner:1987,Schneider-Ehlers-Falco:1992,Erdl-Schneider:1993,Gould:2001} as well as to describe binaries \cite{Congdon-Keeton-book:2018}.  It was recognized that such deviations from spherical symmetry lead to the formation of caustics, which complicates image formation \cite{Ohanian:1983,Blandford-Kovner:1988,Nambu:2013,Chu:2016}. Most of these investigations were conducted using the geometric optics approximation, which is known to be of limited utility when it comes to describing light amplification, especially in the presence of caustics \cite{Gaudi-Petters:2001,Gaudi-Petters:2002}, where such results have singularities. As caustics appear naturally in the point-spread function (PSF) characterizing the optical properties of an extended gravitational lens, there is a need to address these shortcomings.  In particular, it was recognized that a wave-optical treatment of gravitational lensing is needed \cite{Nakamura-Deguchi:1999,Nambu:2012}. Until recently, such a description of an extended lens was not available.

Meanwhile, the solar gravitational lens (SGL) gained attention as a possible means to obtain resolved images of exoplanets \cite{Turyshev:2017,Turyshev-Toth:2017,Turyshev-Toth:2020-im-extend,Toth-Turyshev:2020}. From the beginning, our efforts to describe the SGL were conducted within the Mie theory \cite{Mie:1908,Born-Wolf:1999}, aiming to describe diffraction of electromagnetic (EM) waves by a gravitational field. Such an approach solves a Schr\"odinger-like wave equation for Debye potentials, yielding a wave-optical description for the lens\footnote{In Ref. \cite{Turyshev-Toth:2021-multipoles} we show that although similar results may be obtained within a scalar theory by using a general Fresnel--Kirchhoff diffraction formula or the path integral formalism of quantum field theory, the Mie-inspired solution covers a method to treat vector fields. Such an approach is advantageous from a practical standpoint as it allows us to deal with directly observable quantities and evaluate detection sensitivities (i.e., signal-to-noise ratio \cite{Turyshev-Toth:2020-im-extend,Toth-Turyshev:2020}) while preserving the vectorial nature of the EM field in a weak gravitational field.}. At first, the lens was modeled as a gravitational monopole. This established a good foundation on which increasingly refined models could be constructed. These refinements were needed to capture the fact that the Sun is not a perfect sphere: its rotation and the resulting oblateness result in small axisymmetric perturbations of its otherwise spherically symmetric gravitational field in the form of the quadrupole and, to a lesser extent, higher-order zonal harmonics.

Although our initial objective was to capture only the dominant quadrupole perturbation captured by the $J_2$ zonal harmonics, we were able to do much more. We developed a wave-optical treatment that we call the angular eikonal method, which can be used to describe gravitational lensing by any axisymmetric gravitational field that is dominated by the monopole potential, but perturbed by an infinite set of zonal harmonics \cite{Turyshev-Toth:2021-multipoles}.
The resulting wave-optical treatment of gravitational lensing focuses on evaluating the eikonal phase shift that an EM wave acquires as it travels from the source to the image plane. This phase shift now may be evaluated for any gravitational potential that can be modeled as a perturbed monopole gravitational field. The new method may in fact be used to recover the multipole moments that characterize the mass distribution of the lens, and thus recover its basic geometry and structure.  This wave-optical approach is especially useful to describe imaging with realistic axisymmetric astrophysical lenses \cite{Turyshev-Toth:2021-imaging}.

Our prior work on gravitational lensing focused on the strong interference region of the lens that exists in the vicinity of its primary optical axis (see Fig.~\ref{fig:regions}). As we move further away from that axis,  we enter the weak interference region and then the region of geometric optics (see description in \cite{Turyshev-Toth:2017}.) Clearly, the farther we are from the optical axis, the less is the impact of perturbations to the monopole gravitational field. At some distance from the optical axis, the behavior of an extended gravitational lens becomes indistinguishable from that of a monopole lens. In any case, a complete description of the transition process between various regions is needed to fully understand the behavior of the PSF of the lens. The images seen by a telescope at different distances from the optical axis are also of interest.  A similar discussion in the context of a monopole lens was presented in \cite{Turyshev-Toth:2019-extend,Turyshev-Toth:2020-im-extend}. We can now extend these results to the case of a generic lens that can be described as a perturbed gravitational monopole.

In this paper, we apply our new approach beyond the strong interference region, describing gravitational lensing in all lensing regimes.  This paper is organized as follows:
In Section~\ref{sec:EM-field} we summarize the wave-optical solution we call \emph{the angular eikonal method} \cite{Turyshev-Toth:2021-multipoles} that allows us to determine the EM field in all regions behind the extended axisymmetric SGL, including the regions of strong and weak interference as well as the region of geometric optics.
In Section~\ref{sec:v-large-disp} we address imaging with the SGL of the extended Sun, where we describe the signal received at the focal plane of an imaging telescope which moves in the image plane.
In Section~\ref{sec:sims} we demonstrate the power of our formalism by presenting results that show the view of a point source, projected by the SGL and observed at various distances from the optical axis by an imaging telescope. We also present simulations of an extended source modeling light from a distant resolved star.
In Section~\ref{sec:end} we present our conclusions and identify next steps.
In Appendix~\ref{sec:lim-cases} we consider some limiting cases and demonstrate agreement between previously obtained results and the results in the present paper.

\section{The EM field all regions behind the lens}
\label{sec:EM-field}

In Ref.~\cite{Turyshev-Toth:2021-multipoles}, we studied diffraction of EM waves in the presence of gravity. For this we considered a Mie problem with the electromagnetic field propagating in the vicinity of an extended gravitational lens in the first post-Newtonian approximation of the general theory of relativity. We were able to reduce the problem to a Schr\"odinger-like equation describing the Debye potential and then derived a complete solution for the EM field on an image plane positioned behind the lens. The resulting EM field was used to compute the energy flux in various regions behind that lens by calculating the Poynting vector -- the quantity that is needed to study the optical properties of the lens\footnote{To simplify the material and keep the focus of this paper on its broader objectives, here we only summarize the solution, inviting the reader to consult \cite{Turyshev-Toth:2021-multipoles} for technical details if needed.}.

\subsection{Summary of the solution}

\begin{figure}
\includegraphics[scale=0.27]{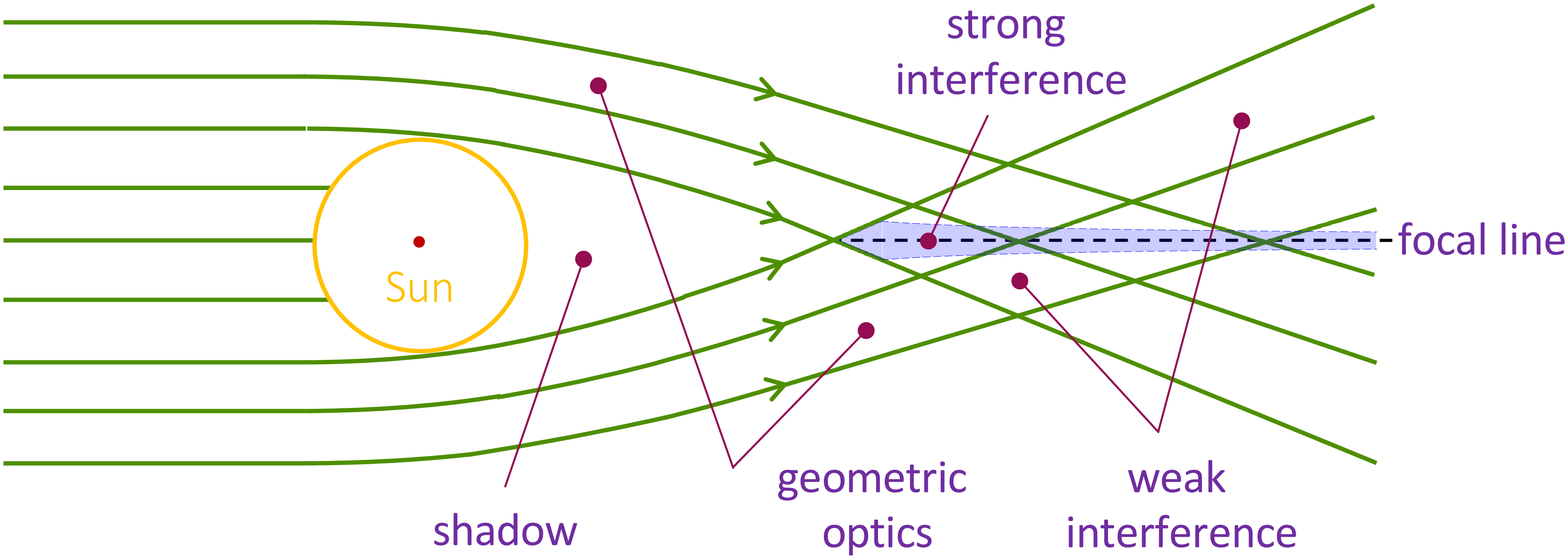}
\caption{\label{fig:regions}The different optical regions of the SGL
(adapted from \cite{Turyshev-Toth:2019-extend}).
}
\end{figure}

We use a heliocentric spherical coordinate system $(r,\theta,\phi)$ and consider a source positioned at a distance $r_0$ from a lens. In \cite{Turyshev-Toth:2021-multipoles}, we studied propagation of a light ray with impact parameter $b$ with respect to the lens and determined the components of the EM field that would be observed on an image plane at distance $r$ from the lens. For a high-frequency EM wave (i.e., neglecting terms $\propto(kr)^{-1}$) and for $r\gg r_g$ (with $r_g=2GM/c^2$ being the Schwarzschild radius of the lens), we derived the EM field  that is needed to estimate the flux through the image plane. Following the logic of solving the Mie problem \cite{Born-Wolf:1999,Turyshev-Toth:2017}, this field can be given to the required order in the following form \cite{Turyshev-Toth:2021-multipoles}:
{}
\begin{eqnarray}
    \left( \begin{aligned}
{D}_\theta& \\
{B}_\theta& \\
  \end{aligned} \right) =    \left( \begin{aligned}
{B}_\phi& \\
-{D}_\phi& \\
  \end{aligned} \right)&=&
 \left( \begin{aligned}
 \cos\phi& \\
 \sin\phi& \\
  \end{aligned} \right)\,e^{-i\omega t}\gamma(r, \theta)+{\cal O}(r_g^2, \theta^2, b/r_0),
  \label{eq:DB-sol-rho_go}
\end{eqnarray}
with the term $\gamma(r, \theta)$ given to ${\cal O}\big({r_g}/{r},r_g^2\big)$ as
{}
\begin{eqnarray}
 \gamma(r,\theta,\phi) &=&  \frac{E_0}{r_0}\frac{ue^{ik(r+r_0+r_g\ln 4k^2rr_0)}}{ikr}\sum_{\ell=kR^\star_\odot}^\infty\frac{\ell+{\textstyle\frac{1}{2}}}{\ell(\ell+1)}e^{i\big(2\sigma_\ell+\frac{\ell(\ell+1)}{2k\tilde r}+2\xi_b
\big)}
\Big\{\frac{\partial P^{(1)}_\ell(\cos\theta)}
{\partial \theta} +\frac{P^{(1)}_\ell(\cos\theta)}{\sin\theta}
 \Big\},
  \label{eq:beta*1*}
\end{eqnarray}
where $1/\tilde r=1/r+1/r_0$ (as discussed in \cite{Turyshev-Toth:2019-extend}) and $\sigma_\ell$ is the Coulomb phase shift (see details in \cite{Turyshev-Toth:2017}). The summation is conducted over the partial momenta $\ell$ that, in a semiclassical analogy, is related to the impact parameter $b$ as $\ell=kb$. Also, the sum in (\ref{eq:beta*1*}) starts at $\ell=kR^\star_\odot$, that corresponds to applying a fully-absolving boundary condition, capturing the fact that light rays with impact parameters $0\leq b < R^\star_\odot=R_\odot+r_g$ are completely absorbed by the opaque Sun.
As usual, $P^{(1)}_\ell(\cos\theta)$ are Legendre polynomials of the first kind \cite{Abramovitz-Stegun:1965}. The radial components of the EM wave behave as $({D}_r, {B}_r)\sim {\cal O}({\rho}/{z},b/r_0)$; thus they are negligibly small compared to the other two components (\ref{eq:DB-sol-rho_go}).

The quantity $\xi_b$  in the phase of (\ref{eq:beta*1*}) is the eikonal phase shift that is acquired by an EM wave as it travels in the vicinity of an extended axisymmetric gravitational lens (such as our Sun). To establish the form of this quantity,  in \cite{Turyshev-Toth:2021-multipoles} we used a heliocentric coordinate system with its $z$-axis aligned with the wave vector $\vec k$ of the incident wave, so that $\vec k=(0,0,1)$, and introduce the vector of the impact parameter, $\vec b=b \vec n_\xi$. Using $z$ to denote the heliocentric distance of the image plane, we define  $\vec x$ to mark a position in the image plane. Lastly, we introduce a unit vector in the direction of the solar rotation axis, $\vec s$. These quantities are given as:
{}
\begin{eqnarray}
{\vec b}&=&b(\cos\phi_\xi,\sin \phi_\xi,0),
\label{eq:note-b}\\
{\vec x}&=&\rho(\cos\phi,\sin \phi,0),
\label{eq:note-x}\\
{\vec s}&=&(\sin\beta_s\cos\phi_s,\sin\beta_s\sin\phi_s,\cos\beta_s).
\label{eq:note}
\end{eqnarray}
With this parametrization, the additional eikonal phase shift $\xi_b$ induced by an extended, axisymmetric and rotating gravitational lens characterized in terms of zonal harmonics was determined \cite{Turyshev-Toth:2021-multipoles} to have the form
{}
\begin{eqnarray}
\xi_b=-kr_g\sum_{n=2}^\infty\frac{J_n}{n}\Big(\frac{R_\odot}{b}\Big)^n \sin^n\beta_s\cos[n(\phi_\xi-\phi_s)],
  \label{eq:xi_b}
\end{eqnarray}
where $J_n$ are the zonal harmonic coefficients of the gravitational field of the lens, such as the SGL.

In \cite{Turyshev-Toth:2021-multipoles}, we considered solution (\ref{eq:DB-sol-rho_go})--(\ref{eq:xi_b}) only in the strong interference region that lies in the proximity of the primary optical axis where $\theta\simeq \sqrt{2r_g/r}$.  Our objective for this paper is to use the solution above and derive results  that will be applicable in all the gravitational lensing regions that are formed behind the else that also include the weak interference region and that of the geometric optics.

\subsection{Eikonal correction for the azimuthal term}

To evaluate expression (\ref{eq:beta*1*}), following \cite{Turyshev-Toth:2021-multipoles},  we use  the asymptotic representation for $P_\ell(\cos\theta)$ and $\ell\gg1$ from\footnote{For an improved, explicit, uniformly valid two-term asymptotic form of this expression, see \cite{Bakaleinikov:2020}} \cite{Bateman-Erdelyi:1953,Korn-Korn:1968,Kerker-book:1969,Abramovitz-Stegun:1965}:
{}
\begin{eqnarray}
P_\ell(\cos\theta)&=& \sqrt{\frac{\theta}{\sin\theta}} J_0\big(\ell \theta\big)+{\cal O}(\theta^2).
\label{eq:Bess0}
\end{eqnarray}

Next, we use expression
{}
\begin{eqnarray}
P^{(1)}_\ell(\cos\theta)=-\frac{\partial P_\ell(\cos\theta)}{\partial\theta}=\ell J_1(\ell\theta)+
{\textstyle\frac{1}{6}}\theta J_0(\ell\theta) +{\cal O}(\theta^2),
\label{eq:Bess}
\end{eqnarray}
alongside with the recurrence relations for the Bessel functions \cite{Abramovitz-Stegun:1965}
{}
\begin{eqnarray}
\frac{2n}{x}J_n(x)=J_{n-1}(x)+J_{n+1}(x),
\label{eq:Bess-rec}
\end{eqnarray}
and derive the following two well-known \cite{Born-Wolf:1999} and  useful relations
{}
\begin{eqnarray}
\frac{P^{(1)}_\ell(\cos\theta)}{\sin\theta}&=& {\textstyle\frac{1}{2}}\ell^2\Big(J_0(\ell \theta)+J_2(\ell \theta)\Big),
\label{eq:pi-l=}
\qquad
\frac{dP^{(1)}_\ell(\cos\theta)}{d\theta}= {\textstyle\frac{1}{2}}\ell^2\Big(J_0(\ell \theta)-J_2(\ell \theta)\Big).
\label{eq:tau-l=}
\end{eqnarray}

Substituting (\ref{eq:pi-l=}) in expression (\ref{eq:beta*1*}), and, following the approach that we presented in \cite{Turyshev-Toth:2019-extend}, we consider the case of the large partial momenta, $\ell\gg1$, which is certainly valid here, as the integration is done from $\ell=kR_\odot\gg1$ to infinity. In this case,  the term $\gamma(r, \theta,\phi)$ is determined from the following integral:
{}
\begin{eqnarray}
\gamma(r,\theta,\phi) =
\frac{E_0}{r_0}\frac{ue^{ik(r+r_0+r_g\ln 4k^2rr_0)}}{ikr}\int_{\ell=kR^\star_\odot}^\infty
\ell d\ell e^{i\big(2\sigma_\ell+\frac{\ell^2}{2k\tilde r}+
2\xi_b\big)}\Big(
J_0(\ell\theta) +{\cal O}\big(\theta^2,\frac{r_g}{r},r_g^2\big)\Big).
  \label{eq:gamma**1*}
\end{eqnarray}

To evaluate this integral we used the \emph{angular eikonal method} presented in \cite{Turyshev-Toth:2021-multipoles}. For that, we first recognize that in the case of a point mass (i.e, when only monopole is present), the resulting gravitational field is spherically symmetric \cite{Turyshev-Toth:2017,Turyshev-Toth:2019}. However, once we include the field from the gravitational multipoles, that symmetry is broken as the eikonal phase shift acquires an azimuthal term, namely $\xi_b=\xi_b(b, \theta,\phi)$. However, we found a way to develop the treatment of the problem even in this generic case.  First, we recall that the eikonal phase shift $\xi_b$ was obtained through an iterative process involving the eikonal approximation that originates from the field of high-energy particle physics but is also applicable in the optics domain.
Next, we recognize that for a spherically symmetric field (which is used as the starting point of our iterative process),  the Bessel function $J_0(\ell \theta)$ can be used in its integral form\footnote{Note that we can use the same representation of this function with the positive sign in the phase, but the result is identical as it will be integrated over the entire range of the azimuthal angle $\phi_\xi$. }
{}
\begin{eqnarray}
J_0(\ell\theta)&=& \frac{1}{2\pi}\int_0^{2\pi} d\phi_\xi e^{-i\ell\theta\cos(\phi_\xi-\phi)}.
\label{eq:J0}
\end{eqnarray}
This is the natural step that captures the spherical symmetry of the field of a gravitational monopole. So, the iterative process used to derive the eikonal phase is conducted under this integral over all the azimuthal angles, $\phi_\xi$. However, in the case of the multipoles the azimuthal symmetry is broken. The presence of this integral over $d\phi_\xi$ allows us to account for this azimuthal dependence within the angular eikonal approximation; hence the name of the method.

We now substitute (\ref{eq:J0}) into  (\ref{eq:gamma**1*}) and see that  expression (\ref{eq:gamma**1*}), to the order of ${\cal O}\big(\theta^2,{r_g}/{r},r_g^2\big)$, transforms as
{}
\begin{eqnarray}
\gamma(r,\theta,\phi) &=&
\frac{E_0}{r_0} \frac{ue^{ik(r+r_0+r_g\ln 4k^2rr_0)}}{ikr} \frac{1}{2\pi}\int_0^{2\pi} d\phi_\xi \int_{\ell=kR^\star_\odot}^\infty
\ell d\ell e^{i\big(2\sigma_\ell+\frac{\ell^2}{2k\tilde r}+2\xi_b-\ell\theta\cos(\phi_\xi-\phi)\big)}.
  \label{eq:beta*3}
\end{eqnarray}

In this form, the integral over $d\phi_\xi$ properly acts not only on the monopole term represented by the term $2\sigma_\ell+\frac{\ell^2}{2k\tilde r}-\ell\theta\cos(\phi_\xi-\phi)$ in the phase of the expression \ref{eq:beta*3}, but on the entire phase  $2\sigma_\ell+\frac{\ell^2}{2k\tilde r}+2\xi_b-\ell\theta\cos(\phi_\xi-\phi)$, which now includes contributions from nonspherical  parts of the gravitational potential via the eikonal phase term, $2\xi_b$. This process constitutes the \emph{angular eikonal method},  valid for weak gravitational fields, which allows us to study the scattering of light on nonspherical potentials under the eikonal approximation.

\subsection{Taking the integral over $b$ with the method of stationary phase}

To develop a solution for (\ref{eq:beta*3}), and for convenience, we use (\ref{eq:xi_b}) and introduce quantity $\psi(\vec b)$, as
{}
\begin{eqnarray}
\xi_b(\vec b)=-kr_g\psi(\vec b),\, \qquad
\psi(\vec b)=\sum_{n=2}^\infty\frac{J_n}{n}\Big(\frac{R_\odot}{b}\Big)^n \sin^n\beta_s\cos[n(\phi_\xi-\phi_s)].
  \label{eq:psi}
\end{eqnarray}

Furthermore, for $\ell\gg kr_g$, evaluate $\sigma_\ell$ as \cite{Turyshev-Toth:2018-grav-shadow}:
{}
\begin{eqnarray}
\sigma_\ell&=& -kr_g\ln \ell.
\label{eq:sig-l*}
\end{eqnarray}
This form agrees with the other known forms of $\sigma_\ell$ \cite{Cody-Hillstrom:1970,Barata:2009ma} that are approximated for large $\ell$ (see discussion in \cite{Turyshev-Toth:2017,Turyshev-Toth:2019-image}).

We rely on the semiclassical approximation (see relevant discussion in \cite{Turyshev-Toth:2017,Turyshev-Toth:2019}) that connects the partial momenta, $\ell$, to the impact parameter $b$:
{}
\begin{equation}
\ell\simeq kb,
\label{eq:S-l-pri-p-g}
\end{equation}
which is applicable for small angles $\theta$ (or, large distances from the Sun, $R_\odot/r<b/r\ll 1$),  (see \cite{Turyshev-Toth:2017} for details) and present the phase in (\ref{eq:beta*3}) as
{}
\begin{eqnarray}
\varphi(\vec b)&=&2\sigma_\ell+\frac{\ell^2}{2k\tilde r}+2\xi_b-\ell\theta\cos(\phi_\xi-\phi)=
k\Big\{\frac{b^2}{2\tilde r} -b\theta\cos(\phi_\xi-\phi)-2r_g\big(\ln kb+\psi(\vec b)\big)\Big\}.
  \label{eq:beta*3+4}
\end{eqnarray}

To establish the nature of this quantity, we recognize from (\ref{eq:note-b}) that the vector of the impact parameter has the form given by (\ref{eq:note-b}) as ${\vec b}=b(\cos\phi_\xi,\sin\phi_\xi,0)$. With this, we define the vector ${\vec \theta}$ to a point on the image plane with coordinates $(r, \theta,\phi)$ that has the form ${\vec \theta}=\theta(\cos\phi,\sin\phi,0)$, developed from (\ref{eq:note-x}) with $\theta=\rho/r$. With these definitions, we see that $b\theta\cos(\phi_\xi-\phi)=({\vec b}\cdot{\vec \theta})$, and, thus (\ref{eq:beta*3+4}) takes the form
{}
\begin{eqnarray}
\varphi(\vec b)&=&
k\Big\{\frac{1}{2\tilde r}\big({\vec b} -\tilde r \vec \theta\big)^2-2r_g\big(\ln kb+\psi(\vec b)\big)\Big\}+{\cal O}(\theta^2).
  \label{eq:beta*3+}
\end{eqnarray}
Thus, $\varphi(\vec b)$ represents the Fermat potential that governs the gravitational lensing phenomena \cite{Liebes:1964,Refsdal:1964,Schneider-Ehlers-Falco:1992}. Note that the same form of the expression (\ref{eq:beta*3+}) may be obtained within the path integral approach (see \cite{Turyshev-Toth:2021-multipoles}).

Using (\ref{eq:beta*3+4}), we can present  the $\gamma(r,\theta,\phi) $ factor from (\ref{eq:beta*3}) as
{}
\begin{eqnarray}
\gamma(r,\theta,\phi) &=&
\frac{E_0}{r_0}ue^{ik(r+r_0+r_g\ln 4k^2rr_0)-i{\textstyle\frac{\pi}{2}}}
\frac{k}{2\pi r}\int_0^{2\pi} d\phi_\xi \int_{b=R^\star_\odot}^\infty
b db \, e^{ik\big(\frac{b^2}{2\tilde r} -b\theta\cos(\phi_\xi-\phi)-2r_g\big(\ln kb+\psi(\vec b)\big)\big)}.
  \label{eq:gamma0}
\end{eqnarray}

We recognize that this is the double integral is with respect to the impact parameter, $\vec b$, specifically, $d^2{\vec b}=d\phi_\xi bdb$. We may try to take this integral with using the 2-dimensional method of stationary phase. However, as even the presence a quadrupole leads to appearance of caustics such a solution will not be precise. This problem can be mitigated if we take only one of the two integrals in (\ref{eq:gamma0}) may be taken with the method of stationary phase. When the multipoles represent a small distortion of the gravitational field, contributions to the eikonal phase shift due to multipoles (\ref{eq:psi}) are much smaller than that of the monopole (given by the $\ln kb$ term in (\ref{eq:beta*3+})). In this case, we may take the integral over $db$ using the method of stationary phase, leaving the integral over $d\phi_\xi$ unevaluated.

In case of gravitation, the monopole term is responsible for a long-range gravitational field that effects rays of light over very large distances (similar to the effect of a Coulomb potential  in the time-independent Schr\"odinger equation). In that case, the behavior of the light ray is well-understood. The impact of the monopole is affects the light ray's trajectory along its entire path from emission to reception.  Within the eikonal approximation, any multipole distortion leads to a phase shift in addition to that produced by the monopole. This makes it possible ot evaluate the radial integral using the method of stationary phase. It can be shown that the error incurred by doing so is of $\sim{\cal O}\big((kr_g)^{-1}\big)$, which is negligible in practice. This justifies our approach in case of a weak gravitational field and its long-range behavior.

\subsubsection{Solving for the impact parameter for the stationary phase}

As was done in \cite{Turyshev-Toth:2019-extend}, we evaluate this integral using the method of stationary phase. To do that, we note that the relevant $b$-dependent part of the phase in (\ref{eq:gamma0})  is of the form (\ref{eq:beta*3+4}).  The phase is stationary when $d\varphi (\vec b)/db=0$, which implies
{}
\begin{equation}
\frac{b}{\tilde r}-\theta\cos(\phi_\xi-\phi)-\frac{2r_g}{b}\Big(1-\sum_{n=2}^\infty J_n\Big(\frac{R_\odot}{b}\Big)^n \sin^n\beta_s\cos[n(\phi_\xi-\phi_s)]\Big)={\cal O}\big(r_g^2, \theta^2\big).
\label{eq:S-l-pri=}
\end{equation}
We solve this equation iteratively, using a trial solution $b=b_{[0]}+b_{[1]}+{\cal O}(J_n^2)$, which allows us to form two equations:
{}
\begin{eqnarray}
b_{[0]}^2- b_{[0]}\tilde r \theta \cos(\phi_\xi-\phi)-2r_g\tilde r&=&{\cal O}\big(r_g^2, \theta^2\big),
\label{eq:b0}\\
b_{[1]}\Big(2b_{[0]}-\tilde r \theta\cos(\phi_\xi-\phi)\Big)+
2r_g\tilde r\sum_{n=2}^\infty J_n\Big(\frac{R_\odot}{b_{[0]}}\Big)^n \sin^n\beta_s\cos[n(\phi_\xi-\phi_s)]&=&{\cal O}\big(r_g^2, \theta^2\big).
\label{eq:b1}
\end{eqnarray}

The quadratic equation (\ref{eq:b0}) yields the following two solutions:
{}
\begin{eqnarray}
b_{[0]}^{\pm}= {\textstyle\frac{1}{2}} \tilde r \theta\cos(\phi_\xi-\phi)\pm\sqrt{\big({\textstyle\frac{1}{2}} \tilde r \theta\cos(\phi_\xi-\phi)\big)^2+2r_g \tilde r}+{\cal O}(\theta^3,r_g^2,J_n).
\label{eq:S-l-pri+0}
\end{eqnarray}
We require the impact parameter to be positive. This condition is be satisfied only for the positive sign in (\ref{eq:S-l-pri+0}). Thus, the  impact parameter $b^{[0]}$ has the form
{}
\begin{eqnarray}
b_{[0]}= \sqrt{\big({\textstyle\frac{1}{2}} \tilde r \theta\cos(\phi_\xi-\phi)\big)^2+2r_g \tilde r}+{\textstyle\frac{1}{2}} \tilde r \theta\cos(\phi_\xi-\phi)+{\cal O}(\theta^3,r_g^2,J_n).
\label{eq:S-l-pri+}
\end{eqnarray}

As this solution has dependence on the azimuthal angle $\phi$, in the case of a monopole, (\ref{eq:S-l-pri+}) actually represents two families of impact parameters when $\phi_\xi-\phi=0$ and $\phi_\xi-\phi=\pi$, yielding
{}
\begin{eqnarray}
b^{[0]}_{\tt in}=\sqrt{({\textstyle\frac{1}{2}} \tilde r \theta)^2+2r_g \tilde r}+{\textstyle\frac{1}{2}} \tilde r \theta, \qquad
b^{[0]}_{\tt sc}=\sqrt{({\textstyle\frac{1}{2}} \tilde r \theta)^2+2r_g \tilde r}-{\textstyle\frac{1}{2}} \tilde r \theta,
\label{eq:S-l-pri0d+}
\end{eqnarray}
where $b^{[0]}_{\tt in}$ and $b^{[0]}_{\tt sc}$ are the two impact parameters describing incident and scattered EM waves, corresponding to light rays passing by the near side and the far side of the Sun (with respect to the location of the telescope), correspondingly. After it is diffracted by a point-source gravitational lens, a wavefront is described as the sum of a gravity-modified plane wave (the incident wave) and a spherical wave centered on the gravitational lensing source (the scattered wave);
see, for instance, Fig.~2 of \cite{Turyshev-Toth:2017}.  The impact parameters $b^{[0]}_{\tt in}$ and $b^{[0]}_{\tt sc}$ correspond to images that appear close to the Einstein ring on opposite sides of the lens; the ``scattered'' image, denoted by ``${\tt sc}$'', on the far side relative to the telescope (called the minor image) always appears inside the Einstein ring, and the ``incident'' image, denoted by ``${\tt in}$'' on the near side always appears outside (major image, see \cite{Schneider-Ehlers-Falco:1992} for details).

For $\sqrt{2r_g/\tilde r}\ll \theta$, this result is equivalent to the two solutions derived  in Sec.~IV of  \cite{Turyshev-Toth:2019-extend}. However, the form (\ref{eq:S-l-pri+}) allows us to study the behavior  of the EM wave in the transition between the two solutions in the region where angle $\theta$ is of the same order as the Einstein deflection angle $\theta\sim \sqrt{2r_g/\tilde r}$.

It is convenient to use a shorthand notation $\theta\cos(\phi_\xi-\phi)=({\vec n}_\xi\cdot{\vec \theta})$, where ${\vec \theta}=\theta(\cos\phi,\sin\phi,0)$. Then, by dividing the solution (\ref{eq:S-l-pri+}) by $\tilde r$, we may present the two solutions in term of the angles $\theta^{[0]}_+=b^{[0]}_{\tt in}/\tilde r$, for $\phi_\xi-\phi=0$, and $\theta^{[0]}_-=b^{[0]}_{\tt in}/\tilde r$, for $\phi_\xi-\phi=\pi$, to ${\cal O}(\theta^3,r_g^2,J_n)$, we have
{}
\begin{equation}
\theta^{[0]}= {\textstyle\frac{1}{2}} \Big(\sqrt{({\vec n}_\xi\cdot{\vec \theta})^2+4\theta_E^2}+({\vec n}_\xi\cdot{\vec \theta})\Big)\quad\rightarrow \quad \theta^{[0]}_+= {\textstyle\frac{1}{2}} \Big(\sqrt{ \theta^2+4\theta_E^2}+ \theta\Big) \qquad {\rm and} \qquad
\theta^{[0]}_-= {\textstyle\frac{1}{2}} \Big(\sqrt{ \theta^2+4\theta_E^2}-\theta\Big),
\label{eq:S-l-pri-mic}
\end{equation}
where $\theta_E=\sqrt{{2r_g}/{\tilde r}}$ is the Einstein deflection angle.  This establishes the correspondence of our analysis in this section to the well-known modeling of microlensing \cite{Liebes:1964,Refsdal:1964,Schneider-Ehlers-Falco:1992}.

Expressions (\ref{eq:S-l-pri-mic})  lead to the familiar expression to describe the image magnification of $A=(u^2+2)/(u\sqrt{u^2+4})$, where $u= \theta/\theta_E$.  Our description allows us to develop the vectorial description of the microlensing phenomena and, besides magnification, it also allows us to describe light amplification. Furthermore, our approach allows further improvement -- it allows to describe deflection on multipoles, where the motion is not loner in one plane, but is a function of all three coordinates $(\tilde r, \theta, \phi)$.

To demonstrate this, we continue with the solution of (\ref{eq:b1}). To solve for $b^{[1]}$, we use $b^{[0]}$ from  (\ref{eq:b0}) and substitute it in  (\ref{eq:b1}) to derive $b^{[1]}$ to ${\cal O}\big(r_g^2, \theta^3\big)$:
{}
\begin{eqnarray}
b_{[1]}=-\frac{
r_g\tilde r}{\sqrt{\big({\textstyle\frac{1}{2}} \tilde r \theta\cos(\phi_\xi-\phi)\big)^2+2r_g \tilde r}}\sum_{n=2}^\infty J_n\frac{R^n_\odot \sin^n\beta_s\cos[n(\phi_\xi-\phi_s)]}{\Big(\sqrt{\big({\textstyle\frac{1}{2}} \tilde r \theta\cos(\phi_\xi-\phi)\big)^2+2r_g \tilde r}+ {\textstyle\frac{1}{2}} \tilde r \theta\cos(\phi_\xi-\phi)\Big)^n}.
\label{eq:b2}
\end{eqnarray}

As a result, using expression (\ref{eq:S-l-pri+}) and (\ref{eq:b2}) in the solution to (\ref{eq:S-l-pri=}) takes the form valid to the order of ${\cal O}\big(r_g^2, \theta^3\big)$:
\begin{eqnarray}
b&=&\sqrt{\big({\textstyle\frac{1}{2}} \tilde r \theta\cos(\phi_\xi-\phi)\big)^2+2r_g \tilde r}+{\textstyle\frac{1}{2}} \tilde r \theta\cos(\phi_\xi-\phi)-\nonumber\\
&-&\frac{
r_g\tilde r}{\sqrt{\big({\textstyle\frac{1}{2}} \tilde r \theta\cos(\phi_\xi-\phi)\big)^2+2r_g \tilde r}}\sum_{n=2}^\infty J_n\frac{R^n_\odot \sin^n\beta_s\cos[n(\phi_\xi-\phi_s)]}{\Big(\sqrt{\big({\textstyle\frac{1}{2}} \tilde r \theta\cos(\phi_\xi-\phi)\big)^2+2r_g \tilde r}+ {\textstyle\frac{1}{2}} \tilde r \theta\cos(\phi_\xi-\phi)\Big)^n}.
\label{eq:b12}
\end{eqnarray}

With this result for the impact parameter $b$, we may now proceed with forming the stationary phase solution.

\subsubsection{Computing expressions needed the stationary phase}

To establish the solution with the method of stationary phase, we also need to compute the second derivative of the phase $\varphi ({\vec b})$ from (\ref{eq:beta*3+4}). With respect to $b$, we have
{}
\begin{eqnarray}
\frac{d^2\varphi}{db^2}=k\Big(\frac{1}{\tilde r} +\frac{2r_g}{b^2}+{\cal O}(J_n)\Big).
\label{eq:S-l2+p}
\end{eqnarray}
Note that we need $\varphi''$ only to the order of ${\cal O}(J_n)$ as in the eikonal approximation we may neglect the influence of the short-range potential  (that depends on the mass multipoles) on the amplitude of the EM wave.

Now, using expression for $b\equiv b_0$ from (\ref{eq:b12}), we have
{}
\begin{eqnarray}
\varphi''(b_0)&=&\dfrac{2k}{\tilde r}\frac{\sqrt{\big({\textstyle\frac{1}{2}} \tilde r \theta\cos(\phi_\xi-\phi)\big)^2+2r_g \tilde r}}{\sqrt{\big({\textstyle\frac{1}{2}} \tilde r \theta\cos(\phi_\xi-\phi)\big)^2+2r_g \tilde r}+ {\textstyle\frac{1}{2}} \tilde r \theta\cos(\phi_\xi-\phi)}+
{\cal O}(r_g^2, \theta^3, J_n),
\label{eq:S-l220-g=}
\end{eqnarray}
which is always positive. Next, we compute $\sqrt{{2\pi}/{\varphi''(b_0)}}$ as
{}
\begin{eqnarray}
\sqrt{\frac{2\pi}{\varphi''(b_0)}}&=&\sqrt{\frac{\pi \tilde r}{k}}\Bigg[1+ \frac{{\textstyle\frac{1}{2}} \tilde r \theta\cos(\phi_\xi-\phi)}{\sqrt{\big({\textstyle\frac{1}{2}} \tilde r \theta\cos(\phi_\xi-\phi)\big)^2+2r_g \tilde r}}\Bigg]^{1/2}
+{\cal O}(r_g^2, \theta^3, J_n).
\label{eq:S-l220-g}
\end{eqnarray}

As a result, the amplitude of the integrand in (\ref{eq:gamma0}), for $b$ from (\ref{eq:b12}), is taking the form
{}
\begin{eqnarray}
A(b_0)\sqrt{\frac{2\pi}{\varphi''(b_0)}}
&=&\frac{k}{rr_0}b\sqrt{\frac{2\pi}{\varphi''}}=\frac{\sqrt{\pi k\tilde r}}{r+r_0}\Bigg[\frac{\Big(\sqrt{\Big({\textstyle\frac{1}{2}}  \theta\cos(\phi_\xi-\phi)\Big)^2+\frac{2r_g}{ \tilde r}}+{\textstyle\frac{1}{2}} \theta\cos(\phi_\xi-\phi)\Big)^3}{\sqrt{\Big({\textstyle\frac{1}{2}}  \theta\cos(\phi_\xi-\phi)\Big)^2+\frac{2r_g}{ \tilde r}}}\Bigg]^{1/2}.~~~~~
\label{eq:apm}
\end{eqnarray}

We are now ready to assemble the stationary phase solution for (\ref{eq:gamma0}), treating the radial part of the double integral.

\subsubsection{Summary of the stationary phase solution}

Finally, we need to compute the stationary phase. For this, we substitute the solution for the impact parameter, $ b$, from  (\ref{eq:b12}) into the expression for the phase, $\varphi(\vec b)$ given by (\ref{eq:beta*3+4}):
{}
\begin{eqnarray}
\varphi(b_{\tt in/sc})&=&
k\bigg\{-{\textstyle\frac{1}{2}}  \tilde r \theta\cos(\phi_\xi-\phi)\Big(\sqrt{\Big({\textstyle\frac{1}{2}} \theta\cos(\phi_\xi-\phi)\Big)^2+\frac{2r_g}{\tilde r}}+{\textstyle\frac{1}{2}} \theta\cos(\phi_\xi-\phi)\Big)-\nonumber\\
&&-\,2r_g\ln \Big(\sqrt{\Big({\textstyle\frac{1}{2}}  \theta\cos(\phi_\xi-\phi)\Big)^2+\frac{2r_g}{\tilde r}}+ {\textstyle\frac{1}{2}}  \theta\cos(\phi_\xi-\phi) \Big)+ r_g-2r_g\ln k\tilde r-\nonumber\\
&&-\,2r_g\sum_{n=2}^\infty \frac{J_n}{n}\frac{R^n_\odot \sin^n\beta_s\cos[n(\phi_\xi-\phi_s)]}{\Big(\sqrt{\big({\textstyle\frac{1}{2}} \tilde r \theta\cos(\phi_\xi-\phi)\big)^2+2r_g \tilde r}+ {\textstyle\frac{1}{2}} \tilde r \theta\cos(\phi_\xi-\phi)\Big)^n}\bigg\}.
  \label{eq:stat-ph}
\end{eqnarray}

As a result, the factor $\gamma(\tilde r,\theta,\phi)$ from (\ref{eq:gamma0}) corresponding to the incident EM wave moving towards the interference region is given in the following form:
{}
\begin{eqnarray}
\gamma(\tilde r,\theta,\phi)&=&
\frac{E_0}{r+r_0}\, e^{ik\big(r+r_0+r_g\ln 4k^2rr_0+r_g-2r_g\ln k\tilde r\big)-i{\textstyle\frac{\pi}{4}}} \frac{1}{2\pi}\int_0^{2\pi} d\phi_\xi \,
a(\tilde r,\theta,\phi)
e^{i\varphi_{\tt }(\tilde r,\theta,\phi)}+{\cal O}(\theta^2, \frac{r_g}{r}\theta^2),
\label{eq:Pi_s_exp4+1pp}\\
a(\tilde r,\theta,\phi)&=&
\sqrt{\pi k\tilde r}\Bigg[\frac{\Big(\sqrt{\Big({\textstyle\frac{1}{2}}  \theta\cos(\phi_\xi-\phi)\Big)^2+\frac{2r_g}{ \tilde r}}+{\textstyle\frac{1}{2}} \theta\cos(\phi_\xi-\phi)\Big)^3}{\sqrt{\Big({\textstyle\frac{1}{2}}  \theta\cos(\phi_\xi-\phi)\Big)^2+\frac{2r_g}{ \tilde r}}}\Bigg]^{1/2},
\label{eq:Pi_s_exp4+1pp2}\\
\varphi(\tilde r,\theta,\phi)
&=&-k\bigg\{{\textstyle\frac{1}{2}}   \tilde r\theta\cos(\phi_\xi-\phi)\Big(\sqrt{\Big({\textstyle\frac{1}{2}} \theta\cos(\phi_\xi-\phi)\Big)^2+\frac{2r_g}{\tilde r}}+{\textstyle\frac{1}{2}}  \theta\cos(\phi_\xi-\phi)\Big)+\nonumber\\
&&\hskip 20pt +\,2r_g\ln \Big(\sqrt{\Big({\textstyle\frac{1}{2}}  \theta\cos(\phi_\xi-\phi)\Big)^2+\frac{2r_g}{\tilde r}}+ {\textstyle\frac{1}{2}}  \theta\cos(\phi_\xi-\phi) \Big)+\nonumber\\
&&\hskip 20pt +\,2r_g\sum_{n=2}^\infty \frac{J_n}{n}\frac{R^n_\odot \sin^n\beta_s\cos[n(\phi_\xi-\phi_s)]}{\Big(\sqrt{\big({\textstyle\frac{1}{2}} \tilde r \theta\cos(\phi_\xi-\phi)\big)^2+2r_g \tilde r}+ {\textstyle\frac{1}{2}} \tilde r \theta\cos(\phi_\xi-\phi)\Big)^n}\bigg\}.
\label{eq:Pi_s_exp4+1pp3}
\end{eqnarray}

Results (\ref{eq:Pi_s_exp4+1pp})--(\ref{eq:Pi_s_exp4+1pp3}) provide all the necessary information for us to compute the components of the EM field in all the regions behind the lens in the case of the weak gravitational field.

\subsection{Deriving the EM field on the image plane}

With the expressions developed above, as a result, the components of the incident and scattered EM field from (\ref{eq:DB-sol-rho_go}) to the order of ${\cal O}\big(r_g^2, \theta^2, b/z_0\big)$ take the form
{}
\begin{eqnarray}
    \left( \begin{aligned}
{D}_\theta& \\
{B}_\theta& \\
  \end{aligned} \right)_{\tt \hskip -2pt }
=    \left( \begin{aligned}
{B}_\phi& \\
-{D}_\phi& \\
  \end{aligned} \right)_{\tt \hskip -2pt }&=&
 \frac{E_0}{r+r_0}e^{i\Omega(t)}
   \frac{1}{2\pi}\int_0^{2\pi} d\phi_\xi \,
  { A}_{\tt } (\tilde r,\theta,\phi)
   \left( \begin{aligned}
 \cos\phi& \\
 \sin\phi& \\
  \end{aligned} \right),~~~~
  \label{eq:DB-sol-in}
\end{eqnarray}
where the phase $\Omega(t)$ is given as
{}
\begin{eqnarray}
\Omega(t)=k\big(r+r_0+r_g\ln 4k^2rr_0+r_g-2r_g\ln k\tilde r\big)-{\textstyle\frac{\pi}{4}}-\omega t,~~~~~
\label{eq:omega}
\end{eqnarray}
and the complex amplitude ${ A}_{\tt }={ A}_{\tt }  \big(\vec x,\phi_\xi\big)$ is given as
{}
\begin{eqnarray}
{ A}_{\tt }(\vec x,\phi_\xi)&=&a_{\tt }(\vec x,\phi_\xi)e^{i \varphi_{\tt }(\vec x,\phi_\xi)}=\nonumber\\
&=&
\sqrt{\pi k\tilde r}\Bigg[\frac{\Big(\sqrt{\Big({\textstyle\frac{1}{2}}  \theta\cos(\phi_\xi-\phi)\Big)^2+\frac{2r_g}{ \tilde r}}+{\textstyle\frac{1}{2}} \theta\cos(\phi_\xi-\phi)\Big)^3}{\sqrt{\Big({\textstyle\frac{1}{2}}  \theta\cos(\phi_\xi-\phi)\Big)^2+\frac{2r_g}{ \tilde r}}}\Bigg]^{1/2}\times\nonumber\\
&&\hskip 20pt\times \exp\bigg[-ik\bigg\{{\textstyle\frac{1}{2}}  \tilde r\theta\cos(\phi_\xi-\phi)\Big(\sqrt{\Big({\textstyle\frac{1}{2}}  \theta\cos(\phi_\xi-\phi)\Big)^2+\frac{2r_g}{\tilde r}}+{\textstyle\frac{1}{2}}  \theta\cos(\phi_\xi-\phi)\Big)+\nonumber\\
&&\hskip 54pt +\,2r_g\ln \Big(\sqrt{\Big({\textstyle\frac{1}{2}}  \theta\cos(\phi_\xi-\phi)\Big)^2+\frac{2r_g}{\tilde r}}+ {\textstyle\frac{1}{2}}  \theta\cos(\phi_\xi-\phi) \Big)+\nonumber\\
&&\hskip 54pt +\,2r_g\sum_{n=2}^\infty \frac{J_n}{n}\frac{R^n_\odot \sin^n\beta_s\cos[n(\phi_\xi-\phi_s)]}{\Big(\sqrt{\big({\textstyle\frac{1}{2}} \tilde r \theta\cos(\phi_\xi-\phi)\big)^2+2r_g \tilde r}+ {\textstyle\frac{1}{2}} \tilde r \theta\cos(\phi_\xi-\phi)\Big)^n}\bigg\}\bigg],
  \label{eq:DB-sol-inA}
\end{eqnarray}
where the radial components of the EM waves behave as $({E}_r, {H}_r)_{\tt \hskip 0pt in/sc} \sim {\cal O}({\rho}/{r},b/r_0)$ and, thus, are negligible for any practical purposes. Note that if $ \theta\gg \sqrt{2r_g/\tilde r}$, results are identical to those reported in \cite{Turyshev-Toth:2019-extend}.

As our interest is the EM field on the image plane, it is convenient to transform these solutions to cylindrical coordinates $(\rho,\phi,z)$, as was done in  \cite{Turyshev-Toth:2017,Turyshev-Toth:2019-extend}. Transforming (\ref{eq:DB-sol-in}),  yields the components of both solutions, to ${\cal O}(r_g^2, \theta^2, b/r_0)$, in the form
{}
\begin{eqnarray}
    \left( \begin{aligned}
{E}_\rho& \\
{H}_\rho& \\
  \end{aligned} \right)_{\tt \hskip -2pt  } =    \left( \begin{aligned}
{H}_\phi& \\
-{E}_\phi& \\
  \end{aligned} \right)_{\tt \hskip -2pt  }&=&
\frac{E_0}{r+r_0}e^{i\Omega(t)}  B \big(\vec x\big)
 \left( \begin{aligned}
 \cos\overline \phi& \\
 \sin\overline \phi& \\
  \end{aligned} \right),
  %  \frac{1}{2\pi}\int_0^{2\pi} d\phi_\xi \, { A}_{\tt }  \big(\tilde r,\theta,\phi\big),~~~~~
  \label{eq:DB-sol-in-cc8=}
\end{eqnarray}
where the $z$-components of the EM waves behave as $({E}_z, {H}_z)_{\tt \hskip 0pt } \sim {\cal O}({\rho}/{z},\sqrt{2r_gz}/z_0)$, and where $\overline\phi$ is the angle that corresponds to the rotated $\overline z$ coordinate axis described in \cite{Turyshev-Toth:2019-extend}. The quantity $ B \big(\vec x\big) $ is  the complex amplitude of the EM field has the following form:
{}
\begin{eqnarray}
 B \big(\vec x\big) =
 \frac{1}{2\pi}\int_0^{2\pi} d\phi_\xi \,  { A}_{\tt }  \big(\vec x,\phi_\xi\big).~~~~
  \label{eq:DB-sol-in-cc=}
\end{eqnarray}
With ${ A}_{\tt }  \big(\vec x,\phi_\xi\big)$ given by (\ref{eq:DB-sol-inA}), the complex amplitude takes the form
\begin{eqnarray}
 B \big(\vec x\big) &=&
\sqrt{\pi k\tilde r}  \,
 \frac{1}{2\pi}\int_0^{2\pi} d\phi_\xi \,
\Bigg[\frac{\Big(\sqrt{\Big({\textstyle\frac{1}{2}}  \theta\cos(\phi_\xi-\phi)\Big)^2+\frac{2r_g}{ \tilde r}}+{\textstyle\frac{1}{2}} \theta\cos(\phi_\xi-\phi)\Big)^3}{\sqrt{\Big({\textstyle\frac{1}{2}}  \theta\cos(\phi_\xi-\phi)\Big)^2+\frac{2r_g}{ \tilde r}}}\Bigg]^{1/2}\times\nonumber\\
&&\hskip 20pt\times \exp\bigg[-ik\bigg\{{\textstyle\frac{1}{2}}  \tilde r\theta\cos(\phi_\xi-\phi)\Big(\sqrt{\Big({\textstyle\frac{1}{2}}  \theta\cos(\phi_\xi-\phi)\Big)^2+\frac{2r_g}{\tilde r}}+{\textstyle\frac{1}{2}}  \theta\cos(\phi_\xi-\phi)\Big)+\nonumber\\
&&\hskip 80pt +\,2r_g\ln \Big(\sqrt{\Big({\textstyle\frac{1}{2}}  \theta\cos(\phi_\xi-\phi)\Big)^2+\frac{2r_g}{\tilde r}}+ {\textstyle\frac{1}{2}}  \theta\cos(\phi_\xi-\phi) \Big)+\nonumber\\
&&\hskip 100pt +\,2r_g\sum_{n=2}^\infty \frac{J_n}{n}\frac{R^n_\odot \sin^n\beta_s\cos[n(\phi_\xi-\phi_s)]}{\Big(\sqrt{\big({\textstyle\frac{1}{2}} \tilde r \theta\cos(\phi_\xi-\phi)\big)^2+2r_g \tilde r}+ {\textstyle\frac{1}{2}} \tilde r \theta\cos(\phi_\xi-\phi)\Big)^n}\bigg\}\bigg].~~~~~
  \label{eq:DB-sol-in-cc+}
\end{eqnarray}

This universal expression for the complex amplitude is valid in all regions of an extended axisymmetric lens, including the geometric optics region, weak and strong interference regions. It represents a powerful result that is now applicable in all these diverse regions with very different gravitational lensing behavior.

We may now evaluate the optical performance of the SGL of the extended Sun by computing its PSF. The PSF characterizes the impulse response of the optical system: it maps light from a point source into the image plane. We can follow the approach used in \cite{Turyshev-Toth:2021-multipoles}, using the result (\ref{eq:DB-sol-in-cc8=})--(\ref{eq:DB-sol-in-cc+}) to compute the energy flux in the image region of the lens.
With overline and brackets denoting time averaging and ensemble averaging, the relevant components of the time-averaged Poynting vector for the EM field in the image volume may be given in the following form (see \cite{Turyshev-Toth:2017,Turyshev-Toth:2019,Turyshev-Toth:2019-extend} for details):
 {}
\begin{eqnarray}
S_z({\vec x})=\frac{c}{4\pi}\big<\overline{[{\rm Re}{\vec E}\times{\rm Re}{\vec H}]}_z\big>=\frac{c}{4\pi}\frac{E_0^2}{(r+r_0)^2}
\big<\overline{\big({\rm Re}\big[{ B}({\vec x})e^{i\Omega (t)}\big]\big)^2}\big>=\frac{c}{8\pi}\frac{E_0^2}{(r+r_0)^2}
|B({\vec x})|^2,
  \label{eq:S_z*6z}
\end{eqnarray}
where $|B({\vec x})|^2=B({\vec x})B^*({\vec x})$, with $B^*(\vec x)$ being the complex conjugate of $B({\vec x})$. Note that ${\bar S}_\rho= {\bar S}_\phi=0$ for all practical purposes. Defining light amplification as usual \cite{Turyshev-Toth:2017,Turyshev-Toth:2019,Turyshev-Toth:2019-extend}, $\mu_z({\vec x})=S_z({\vec x})/|\vec S_0({\vec x})|$, where $\vec S_0({\vec x})=(c/8\pi){E_0^2}/{(r+r_0)^2}\, \vec k$ being the Poynting vector carried by a plane wave in the vacuum in flat spacetime, we have the light amplification factor of the lens that, for short wavelengths (i.e., $kr_g\gg1$) is given by
 {}
\begin{eqnarray}
\mu_z({\vec x})=|B({\vec x})|^2.
  \label{eq:S_mu}
\end{eqnarray}
 We recognize that the quantity $\mu_z({\vec x})$ is the PSF of the SGL that is scaled by the amplification factor and it describes all lensing regimes with this extended lens. In Appendix~\ref{sec:small-dev} we show that, in some cases, the amplification factor explicitly multiplies the PSF, but, in general, the PSF (\ref{eq:S_mu}) is implicitly scaled by the amplification factor via (\ref{eq:DB-sol-in-cc+}).

In Appendix~\ref{sec:lim-cases} we consider limiting cases of  $B\big(\vec x) $ from  (\ref{eq:DB-sol-in-cc+}). Those cases include very small deviations $\rho$ from the optical axis, namely $\rho/r\equiv \theta\ll \sqrt{2r_g/r}$; very large deviations $\rho/r\equiv \theta\gg \sqrt{2r_g/r}$; and those in-between. We show that far from the optical axis, the PSF that is constructed with the help of $B({\vec x})$ from (\ref{eq:DB-sol-in-cc+}) exhibits the behavior of the monopole PSF \cite{Turyshev-Toth:2017}, but as we come closer to the optical axis, the effect of multipoles becomes more pronounced, ultimately bringing us to the caustic region, discussed in  \cite{Turyshev-Toth:2021-multipoles}.

\section{Imaging with the  SGL of the extended Sun}
\label{sec:v-large-disp}

The complex amplitude (\ref{eq:DB-sol-in-cc+}) developed in the previous section describes the EM field in the image plane. This field, however, is not what is usually observed. Rather, observations are made with an imaging telescope looking back in the direction of the lens. Our formalism also grants us the ability to accurately describe the image that forms in the focal plane of such a telescope: i.e., the actual observable.

 \subsection{Description of the imaging geometry}
\label{sec:geom}

 With the knowledge of the EM field in the image plane behind an extended gravitational lens  (\ref{eq:DB-sol-in-cc8=})--(\ref{eq:DB-sol-in-cc+}) and following the approach developed in \cite{Turyshev-Toth:2019-image,Turyshev-Toth:2020-im-extend,Turyshev-Toth:2021-imaging}, we can now describe what an imaging telescope would detect on its focal plane.  Such telescopic capability is important as it characterizes the measured optical signal \cite{Turyshev-Toth:2021-imaging,Turyshev-Toth:2021-quartic}.

Similarly to \cite{Turyshev-Toth:2019-image,Turyshev-Toth:2021-imaging}, we describe the geometry of the observation using ${\vec x}$ to represent the current position of an optical telescope in the SGL's image plane, ${\vec x}'$, denoting any point in the same plane, and ${\vec x}_i$, representing a point on the focal plane of the optical telescope. These positions are given as
{}
\begin{eqnarray}
\{{\vec x}\}&\equiv& (x,y,0)=\rho\,\big(\cos\phi,\sin\phi,0\big)=\rho{\vec n}, \label{eq:coord'}\\
\{{\vec x'}\}&\equiv& (x',y',0)=\rho'\big(\cos\phi',\sin\phi',0\big)=\rho'{\vec n'},
\label{eq:x-im} \\
 \{{\vec x}_i\}&\equiv& (x_i,y_i,0)=\rho_i\big(\cos\phi_i,\sin\phi_i,0\big)=\rho_i{\vec n}_i.
  \label{eq:coord}
\end{eqnarray}

To convolve the PSF of the SGL with that of a thin lens that represents an aperture of a telescope, we first need to establish an appropriate form of the PSF for point sources. Examining (\ref{eq:DB-sol-inA}), we see that it contains the expression $\rho\cos(\phi_\xi-\phi)$, which may be transformed as
{}
\begin{eqnarray}
\rho\cos(\phi_\xi-\phi)=({\vec n}_\xi\cdot{\vec x}).
\label{eq:rn}
\end{eqnarray}

We now transition from the current position $\vec x$ of the telescope to an arbitrary location within the telescope's aperture by the substitution
{}
\begin{eqnarray}
\vec x \qquad \Rightarrow \qquad {\vec x}+\vec x'.
\label{eq:rn0}
\end{eqnarray}
Therefore, we may write
{}
\begin{eqnarray}
({\vec n}_\xi\cdot{\vec x}) \qquad \rightarrow \qquad
({\vec n}_\xi\cdot{\vec x}) +({\vec n}_\xi\cdot{\vec x}').
\label{eq:rn2}
\end{eqnarray}
We note that $\vec x'$ varies only with the aperture, whereas  $\vec x$ can be anywhere in the SGL image plane. In much of the image plane, we have  $\rho' \ll \rho$. This allows us to expand (\ref{eq:DB-sol-inA}) in terms of the small parameter  $\rho'/\rho$, keeping only terms of the first order in $\rho'/\rho$. In addition, we recognize that the vector $\vec \theta=\theta(\cos\phi,\sin\phi,0)=\vec x/r$, with $r$ being the distance to the image plane,  may be transformed as
{}
\begin{eqnarray}
({\vec n}_\xi\cdot{\vec \theta})=({\vec n}_\xi\cdot{\vec x})/r \qquad \rightarrow \qquad
({\vec n}_\xi\cdot{\vec x})/r +({\vec n}_\xi\cdot{\vec x}')/r=\frac{\rho}{r}\cos(\phi_\xi-\phi)+\frac{\rho'}{r}\cos(\phi_\xi-\phi').
\label{eq:the}
\end{eqnarray}

This approximation yields the following result for the complex amplitude, ${ A}_{\tt }({\vec x})$, from (\ref{eq:DB-sol-inA}), but given with the shifted argument according to (\ref{eq:rn0}):
{}
\begin{eqnarray}
{ A}_{\tt }({\vec x},{\vec x}')&=&
a_{\tt }(\vec x,\vec n_\xi)
\exp\Big[i\Big(\delta\varphi_{\tt }(\vec x,{\vec n}_\xi)-\nu{\tt }({\vec n}_\xi\cdot{\vec x'})\Big)\Big],
  \label{eq:amp-Ain}
\end{eqnarray}
with the amplitude factor $a_{\tt }(\vec x)$ and phase $\delta\varphi_{\tt }(\vec x) $ given as
{}
  \begin{eqnarray}
a_{\tt }(\vec x,\vec n_\xi) &=& \sqrt{\pi k\tilde r}\Bigg[\frac{\Big(\sqrt{\big({\textstyle\frac{1}{2}} ({\vec n}_\xi\cdot{\vec x})/r \big)^2+\frac{2r_g}{ \tilde r}}+{\textstyle\frac{1}{2}}  ({\vec n}_\xi\cdot{\vec x})/r \Big)^3}{\sqrt{\big({\textstyle\frac{1}{2}}  ({\vec n}_\xi\cdot{\vec x})/r \big)^2+\frac{2r_g}{ \tilde r}}}\Bigg]^{1/2}+{\cal O}(\vec x/r),
    \label{eq:q_insc}\\
    \delta\varphi_{\tt }
(\vec x,\vec n_\xi) &=& -k\bigg\{{\textstyle\frac{1}{2}}  ({\vec n}_\xi\cdot{\vec x}) \Big( \sqrt{\big({\textstyle\frac{1}{2}}  ({\vec n}_\xi\cdot{\vec x})/r \big)^2+\frac{2r_g}{\tilde r}}+{\textstyle\frac{1}{2}}  ({\vec n}_\xi\cdot{\vec x})/r\Big)+
\nonumber\\
&+&
2r_g\bigg(\ln \Big(\sqrt{\big({\textstyle\frac{1}{2}}  ({\vec n}_\xi\cdot{\vec x})/r\big)^2+\frac{2r_g}{\tilde r}}+ {\textstyle\frac{1}{2}}  ({\vec n}_\xi\cdot{\vec x})/r\Big)+
\sum_{n=2}^\infty  \frac{J_n}{n}\frac{R^n_\odot \sin^n\beta_s\cos[n(\phi_\xi-\phi_s)]}{\Big(\sqrt{\big({\textstyle\frac{1}{2}} ({\vec n}_\xi\cdot{\vec x})\big)^2+2r_g \tilde r}+ {\textstyle\frac{1}{2}} ({\vec n}_\xi\cdot{\vec x})\Big)^n}\bigg)\bigg\}.~~~~~
    \label{eq:Ain-d_ph}
\end{eqnarray}
We note that when the angle $\theta$ is large,   $\theta \gg \sqrt{2r_g/\tilde r}$ and thus, $\rho\gg \sqrt{2r_g\tilde r}$, and we get back the PSF of a monopole. Thus, the integral (\ref{eq:beta*3}) may be taken using the method of stationary phase applied to the double integral. In that case, the factors $a_{\tt in/sc}$ in (\ref{eq:q_insc})  take their known values (see \cite{Turyshev-Toth:2019-extend} for details), namely $a^2_{\tt in}(\rho,\tilde r)=1+{\cal O}(r_g\theta^2,r_g^2)$ and $a^2_{\tt sc}(\rho,\tilde r) =({2r_g\tilde r}/{\rho^2})^2 +{\cal O}(r_g\theta^2,r_g^2)$. However, our new expressions (\ref{eq:q_insc}) allow studying the cases when  $\rho\simeq \sqrt{2r_g\tilde r}$ anywhere in the image plane.
The last quantity present in  (\ref{eq:amp-Ain})  is the  spatial frequency $\nu=\nu_{\tt }(\vec x,\vec n_\xi)$, defined as
{}
  \begin{eqnarray}
\nu_{\tt }(\vec x,\vec n_\xi)&=&k\Big(\sqrt{\big({\textstyle\frac{1}{2}}  ({\vec n}_\xi\cdot{\vec x})/r\big)^2+\frac{2r_g}{\tilde r}}+ {\textstyle\frac{1}{2}}  ({\vec n}_\xi\cdot{\vec x})/r\Big).
  \label{eq:betapm}
  \end{eqnarray}

The quantities (\ref{eq:amp-Ain})--(\ref{eq:betapm}) describe the complex amplitude of the EM field, ${B}({\vec x},{\vec x}')$ from (\ref{eq:DB-sol-in-cc=}), as measured in the focal plane of an imaging telescope.

 \subsection{The EM field in the telescope's focal plane}
\label{sec:im-tele}

The focal plane of the optical telescope is located at the focal distance $f$ of the lens, centered on ${\vec x}'$. Using the Fresnel--Kirchhoff diffraction formula, the amplitude of the image field in the optical telescope's focal plane at a location ${\vec x}_i=(x_i,y_i)$ is derived from (\ref{eq:DB-sol-in-cc=}) and is given by  \cite{Wolf-Gabor:1959,Richards-Wolf:1959,Born-Wolf:1999}:
{}
\begin{eqnarray}
{B}({\vec x},{\vec x}_i)=\frac{i}{\lambda}\iint \displaylimits_{|{\vec x}'|^2\leq (d/2)^2} \hskip -7pt  B({\vec x},{\vec x}')e^{-i\frac{k}{2f}|{\vec x}'|^2}\frac{e^{iks'}}{s'}d^2{\vec x}'.
  \label{eq:amp-w-f0}
\end{eqnarray}

The function $\exp[-i\frac{k}{2f}|{\vec x}'|^2]=\exp[-i\frac{k}{2f}(x'^2+y'^2)]$ in (\ref{eq:amp-w-f0}) represents the action of the convex lens that transforms incident plane waves to spherical waves, focusing at the focal point. Assuming that the focal length is sufficiently greater than the radius of the lens, we may approximate the optical path $s'$ as $s'=\sqrt{(x'-x_i)^2+(y'-y_i)^2+f^2}\sim f+\big((x'-x_i)^2+(y'-y_i)^2\big)/2f$. This allows us to present (\ref{eq:amp-w-f0}) as
{}
\begin{eqnarray}
{B}({\vec x},{\vec x}_i)&=&
- \frac{e^{ikf(1+{{\vec x}_i^2}/{2f^2})}}{i\lambda f}\iint\displaylimits_{|{\vec x}'|^2\leq (\frac{1}{2}d)^2} d^2{\vec x}'
 B({\vec x},{\vec x}') e^{-i\frac{k}{f}({\vec x}'\cdot{\vec x}_i)}.
  \label{eq:amp-w-f+}
\end{eqnarray}

Expressions (\ref{eq:amp-Ain})--(\ref{eq:betapm})  allow us to consider imaging of point sources with the SGL, now treated as that produced by a gravitating body that is axisymmetric and rotating, thus admitting characterization of its external gravitational field by zonal harmonics. To accomplish this, following \cite{Turyshev-Toth:2019-image,Turyshev-Toth:2020-im-extend}, we use the expression for $A({\vec x},{\vec x}')$ from (\ref{eq:amp-Ain}) and present the Fresnel--Kirchhoff diffraction formula as
{}
\begin{eqnarray}
{\cal A}_{\tt }({\vec x},{\vec x}_i)&=&
- \frac{e^{ikf(1+{{\vec x}_i^2}/{2f^2})}}{i\lambda f}\iint\displaylimits_{|{\vec x}'|^2\leq (\frac{1}{2}d)^2} d^2{\vec x'}\,
{ A}_{\tt }({\vec x},{\vec x}') e^{-i\eta_i({\vec n}_i\cdot{\vec x}')}=\nonumber\\
&=&- \frac{e^{ikf(1+{{\vec x}_i^2}/{2f^2})}}{i\lambda f}
 a_{\tt}(\vec x,\vec n_\xi)
e^{i\delta\varphi_{\tt }(\vec x,\vec n_\xi)}\,
\iint\displaylimits_{|{\vec x}'|^2\leq (d/2)^2}\hskip -8pt
  d^2{\vec x}'\,
e^{i\big(-\nu_{\tt }({\vec n}_\xi\cdot{\vec x}')-\eta_i({\vec n}_i\cdot{\vec x}')\big)},
  \label{eq:amp-w-f}
\end{eqnarray}
where the spatial frequency $\nu=\nu_{\tt }(\vec x,\vec n_\xi)$ is given by (\ref{eq:betapm}).  Also, for a telescope with focal length of $f$ and for a radial pixel position $\rho_i$, the factor $ \eta_i$ has the form \cite{Turyshev-Toth:2019-image,Turyshev-Toth:2020-im-extend,Turyshev-Toth:2021-imaging}
{}
\begin{eqnarray}
 \eta_i=k\frac{\rho_i}{f}.
\label{eq:zerJ}
\end{eqnarray}

Therefore, to derive  the amplitudes of the EM field in the focal plane of the optical telescope, corresponding to  (\ref{eq:amp-Ain}), we need to evaluate an integral of the type
{}
\begin{eqnarray}
\iint\displaylimits_{|{\vec x}'|^2\leq (d/2)^2}\hskip -8pt
  d^2{\vec x}'\,e^{i\big(- \nu_{\tt }({\vec n}_\xi\cdot{\vec x}')-\eta_i({\vec n}_i\cdot{\vec x}')\big)}.
  \label{eq:amp-int}
\end{eqnarray}
To evaluate this integral, we present the phase  in (\ref{eq:amp-int}) as
{}
\begin{eqnarray}
- \nu_{\tt }({\vec n}_\xi\cdot{\vec x}')-\eta_i({\vec n}_i\cdot{\vec x}')=-u\,\rho' \cos\big(\phi'-\sigma\big)+{\cal O}(\rho^2),
  \label{eq:ph4s}
\end{eqnarray}
where, for convenience, we defined
{}
\begin{eqnarray}
u&=&\sqrt{\nu^2+2\nu\eta_i\cos\big(\phi_\xi-\phi_i\big)+\eta_i^2},
\qquad
\cos\sigma=
\frac{\nu  \cos\phi_\xi+\eta_i\cos\phi_i}{u},
\qquad
\sin\sigma=\frac{\nu \sin\phi_\xi+\eta_i\sin\phi_i}{u}.
\label{eq:vpms}
\end{eqnarray}

With these definitions, and using the parameterization given in (\ref{eq:x-im}), the integral (\ref{eq:amp-int}) may be evaluated as
{}
\begin{eqnarray}
  \int_0^{2\pi} \hskip -4pt d\phi' \int_0^{d/2} \hskip -4pt \rho' d\rho'\,
  e^{-iu\rho'\cos(\phi'-\sigma)}=\pi\Big(\frac{d}{2}\Big)^2\, \frac{
2J_1(u\frac{1}{2}d)}{u\frac{1}{2}d}.
  \label{eq:amp-int*}
\end{eqnarray}
As a result,  using (\ref{eq:amp-Ain}) in  (\ref{eq:amp-w-f})  leads to the following amplitude of the  EM wave on the optical telescope's image plane:
{}
\begin{eqnarray}
{\cal A}_{\tt }({\vec x},{\vec x}_i,\vec n_\xi)&=&
\Big(\frac{kd^2}{8f}\Big)\, \Big\{
a_{\tt }
\Big(\frac{
2J_1(u\frac{1}{2}d)}{u\frac{1}{2}d}\Big)e^{i\big(kf(1+{{\vec x}_i^2}/{2f^2})+\delta\varphi_{\tt }
(\vec x,\vec n_\xi) +\frac{\pi}{2}\big)}+{\cal O}(r_g^2)\Big\}.
  \label{eq:amp-Aind}
\end{eqnarray}

Therefore, the Fourier-transformed complex amplitude (\ref{eq:amp-w-f0}) takes the from
{}
\begin{eqnarray}
&&  \frac{1}{2\pi}\int_0^{2\pi} d\phi_\xi \, {\cal A}_{\tt }({\vec x},{\vec x}_i,\vec n_\xi)=
 \Big(\frac{kd^2}{8f}\Big)e^{i\big(kf(1+{{\vec x}_i^2}/{2f^2}) +\frac{\pi}{2}\big)}
 {\cal B} \big(\vec x,\vec x_i\big),~~~~~
  \label{eq:FtB}
\end{eqnarray}
where ${\cal B} \big(\vec x,\vec x_i\big) $ is given as
{}
\begin{eqnarray}
 {\cal B} \big(\vec x,\vec x_i\big) &=&
 \frac{1}{2\pi}\int_0^{2\pi} d\phi_\xi \,  \Big\{
  a_{\tt }(\vec x,\vec n_\xi)
  \Big(\frac{
2J_1(u({\vec x}_i,\vec x,\vec n_\xi)\frac{1}{2}d)}{u({\vec x}_i,\vec x,\vec n_\xi) \frac{1}{2}d}\Big)
e^{i\delta\varphi_{\tt }(\vec x,\vec n_\xi)}\Big\},~~~~~
  \label{eq:Binsc}
\end{eqnarray}
where $a_{\tt }(\vec x,\vec n_\xi)$,  $\delta\varphi_{\tt }(\vec x,\vec n_\xi)$, and $u({\vec x}_i,\vec x,\vec n_\xi)$ are given by (\ref{eq:q_insc}), (\ref{eq:Ain-d_ph}), and (\ref{eq:vpms}), correspondingly.

Using this result together with (\ref{eq:DB-sol-in-cc8=}), we obtain the EM field on the detector that is given as below
{}
\begin{eqnarray}
    \left( \begin{aligned}
{E}_\rho& \\
{H}_\rho& \\
  \end{aligned} \right)_{\tt} =    \left( \begin{aligned}
{H}_\phi& \\
-{E}_\phi& \\
  \end{aligned} \right)_{\tt}&=&
\frac{E_0}{r+r_0}
 e^{i\big(\Omega(t)+\frac{\pi}{2}+kf(1+{{\vec x}_i^2}/{2f^2})\big)}
  \Big(\frac{kd^2}{8f}\Big)
 {\cal B} \big(\vec x,\vec x_i\big)
 \bigg( \begin{aligned}
\cos\overline \phi& \\
\sin\overline \phi& \\
  \end{aligned} \bigg).
  \label{eq:EM-det}
\end{eqnarray}

After time averaging, we derive the Poynting vector of the EM wave in the focal plane of the imaging telescope:
 {}
\begin{eqnarray}
S_{\tt }({\vec x},{\vec x}_i)&=&\frac{c}{8\pi}
\frac{E_0^2}{(r+r_0)^2}
 \Big(\frac{kd^2}{8f}\Big)^2
  {\cal B}^2 \big(\vec x,\vec x_i\big) .
  \label{eq:PV}
\end{eqnarray}

As a result, the  intensity on the focal plane, ${\cal I}({\vec x},{\vec x}_i)$,  of the system that includes the SGL and a thin lens is given in the form as below:
 {}
\begin{eqnarray}
{\cal I}({\vec x},{\vec x}_i)&=&
  {\cal B}^2 \big(\vec x,\vec x_i\big),
  \label{eq:PFT}
\end{eqnarray}
where the Fourier-transformed complex amplitude $ {\cal B}\big(\vec x,\vec x_i\big)$ from (\ref{eq:Binsc}).
We emphasize that $\mu_z({\vec x})$ from (\ref{eq:S_mu}) is the PSF of the extended SGL. It describes the image of a point source projected on the image plane at the SGL's focal region. At the same time, the quantity ${\cal I}({\vec x},{\vec x}_i)$ from (\ref{eq:PFT}) is the intensity of light received on the focal plane of an imaging telescope. This is a directly observable quantity that is accessible to an optical telescope. As such, it is of most importance for any practical applications of  the SGL.
The resulted expression for the intensity on the focal plane allows considering imaging of various sources with the SGL of an extended Sun. We will do that next.

\section{Application of results}
\label{sec:sims}

\begin{figure}
\includegraphics{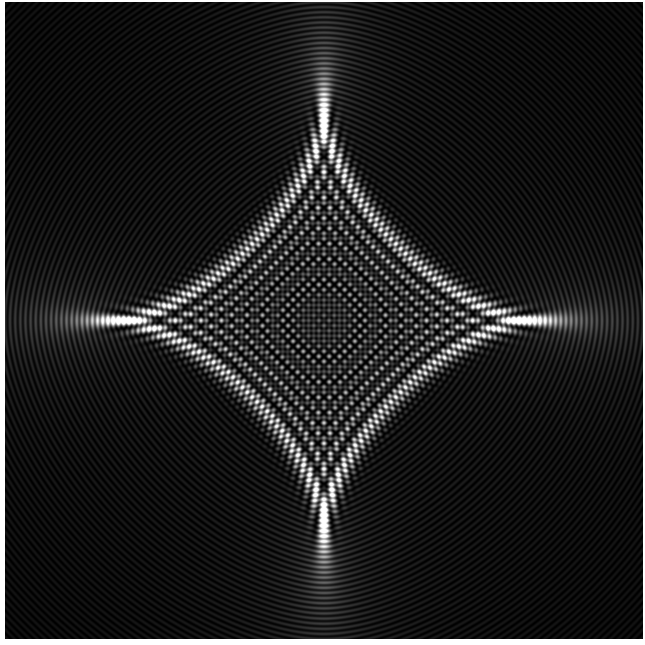}
\caption{\label{fig:astroid}An illustrative example of the SGL PSF, appearing as the astroid caustic projected into the image plane by the SGL, with its recognizable cusps (vertices) and folds. As an imaging telescope enters this region in the image plane, its view of a distant source transitions from a pair of images (the primary and secondary image) into some variation of an Einstein cross or Einstein ring, depending on the size of the astroid, the imaging wavelength, and the size of the light source. Adapted from \cite{Turyshev-Toth:2021-multipoles}.}
\end{figure}

The formalism developed in the preceding section opens the route to simulate the effects of the SGL beyond the immediate vicinity of the optical axis in its strong interference region (see Fig.~\ref{fig:regions}). There is, however, first our obstacle: evaluation of the remaining integral in our final expression (\ref{eq:Binsc}).

\subsection{Evaluation method}

Equation~(\ref{eq:Binsc}) describes the view seen by an imaging telescope of a distant source, both near and far from the optical axis of the gravitational lens. To use this equation, it is necessary to evaluate the remaining integral in the regions of interest. Examining it more closely, we note that the integral has finite integration limits, which makes numerical evaluation easier. However, it is still an oscillatory integral. Moreover, at large distances from the optical axis, the oscillations become very rapid. This makes direct numerical evaluation challenging.

On the other hand, a rapidly oscillating integral implies the possible use of the method of stationary phase once again. This is precisely what we have accomplished in \cite{Turyshev-Toth:2021-quartic}, for the case when $J_4$ and higher order zonal harmonics can be safely neglected, thus leaving only the astroid caustic due to $J_2$. The result, expressed through the roots of a quartic equation, works reliably everywhere in the region of strong interference, only showing occasional rounding errors in the immediate vicinity of the caustic boundary of the projected astroid pattern of a quadrupole lens (Fig.~\ref{fig:astroid} \cite{Turyshev-Toth:2021-multipoles}).

Beyond the region of strong interference, the contribution of the zonal harmonics is negligible and we can use previously developed monopole solutions for efficient evaluation.

Using this combination of methods, we are now in the position to evaluate (\ref{eq:Binsc}) everywhere in the image plane, constructing simulated views of point sources as seen by an imaging telescope through the SGL.

\subsection{Simulated approach to the optical axis}

\begin{figure}
\includegraphics{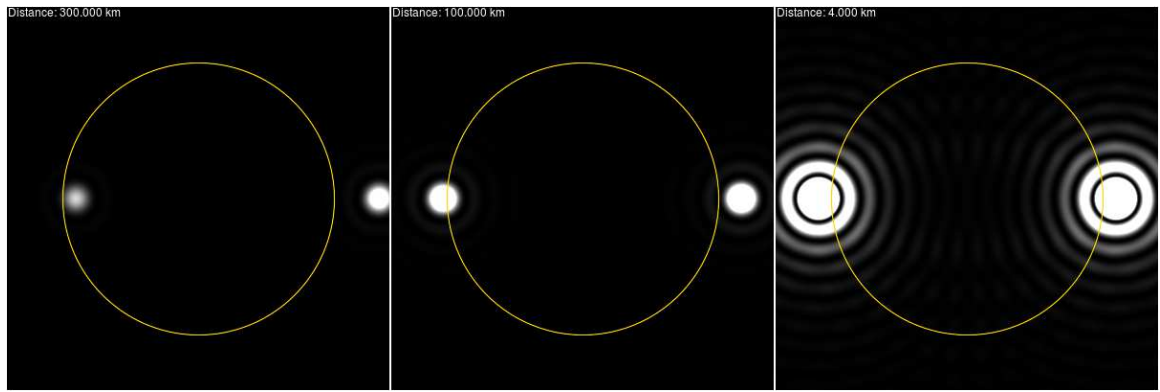}
\caption{\label{fig:appr-pre}View of a distant compact source by a telescope approaching the SGL optical axis associated with that source. The telescope is positioned at $3\times 10^5$~km, $1\times 10^5$~km and $4\times 10^3$~km from the optical axis. Note that at $3\times 10^5$~km, the secondary image of the source is still obscured by the solar disk (shown as a yellow circle). By the time we reach $4\times 10^3$~km, the images become indistinguishable, even as light amplification increases nearly hundredfold. For the full animation, see \protect\url{https://www.vttoth.com/CMS/physics-notes/361}.}
\end{figure}

\begin{figure}
\includegraphics{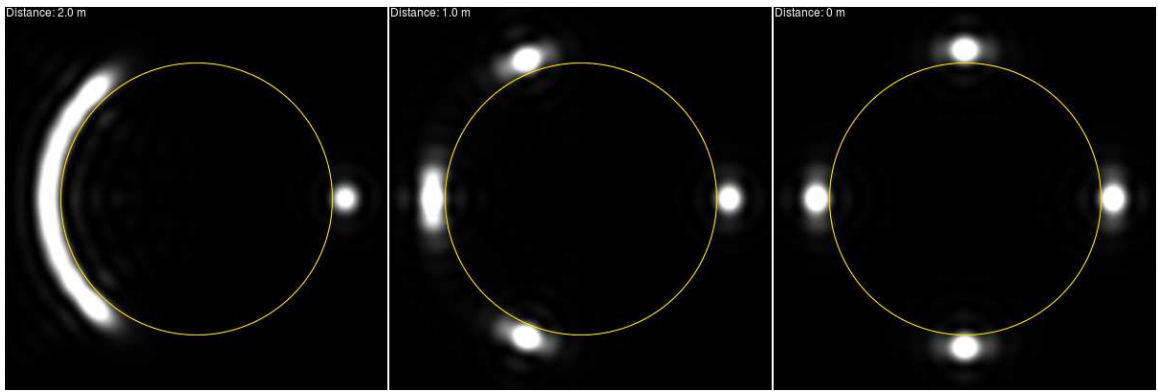}
\caption{\label{fig:approach1}View of a distant point source by a telescope near the optical axis, at 2~m, 1~m and positioned on the axis. The optical axis is at 5.74$^\circ$ from the solar axis of rotation, a direction chosen because it representatively shows the development of an Einstein cross during this final approach. The view is that of a telescope with a 1~m aperture; light amplification is of ${\cal O}(10^8)$.
For the full animation, see \protect\url{https://www.vttoth.com/CMS/physics-notes/361}.}
\end{figure}

To demonstrate the power of the approach captured by the expression (\ref{eq:Binsc}), we chose to simulate the view of a distant point source, as seen by an imaging telescope that is approaches the optical axis of that star with respect to the SGL.

We were able to assemble a series of still images, ultimately in the form of animations\footnote{See \protect\url{https://www.vttoth.com/CMS/physics-notes/361} for a full set of animations.}, which show how an imaging telescope would see the distant source as it was approaches the optical axis that corresponds to that source. Select frames from this animation are presented in this section.

We began the simulation with the imaging telescope located at $10^6$ km from the optical axis, looking in the direction of the Sun (see Fig.~\ref{fig:appr-pre}). This distance was chosen because it is comparable in magnitude to the solar radius, thus placing the imaging telescope firmly in the region of geometric optics.

At the beginning, the source's ``primary image'' is outside the telescope's field of view, and no noticeable ``secondary image'' forms yet on the opposite side of the Sun. At $6\times 10^5$~km from the optical axis, a faint secondary image emerges, or rather, would emerge if the Sun were transparent. In reality, light from that secondary image is yet blocked by the opaque disk of the Sun. When the telescope is at $3\times 10^5$~km from the optical axis (less than half the solar radius) the primary image becomes clearly visible within the imaging area. This is the unobstructed view of the distant source, already amplified by the SGL, so its peak central brightness is $\sim$1.8 times the brightness of the unamplified image. The secondary image, now less faint, is still obscured by the solar disk.

When the telescope is only $\sim1\times 10^5$~km from the optical axis, the secondary image emerges from behind the Sun. Light amplification is becoming significant: the primary image's peak brightness is now more than four times as bright as the unamplified star. When the telescope approaches within $\sim 2\times 10^4$~km of the optical axis, the primary and secondary images are already nearly identical in appearance, at symmetric positions, settling at a distance from the solar limb that corresponds to the radius of a yet-to-form Einstein ring. Light amplification is substantial: the peak brightness that the imaging telescope sees is nearly 20 times the intensity of light from the unamplified star. Even so, the images remain point-like in appearance: This is dictated by the diffraction-limited resolution of the imaging telescope itself.

At this stage, the position of the two images of the point source is final. As the telescope continues to approach the optical axis, however, light amplification increases across several orders of magnitude.

For the purposes of this simulation, we chose to place the optical axis very near the solar axis of rotation, in order to keep the contribution of the $J_2$ zonal harmonic small.  Figure \ref{fig:approach1} shows the telescope's final approach to an optical axis that is at $5.74^\circ$ from the solar axis of rotation, which corresponds to $\sin\beta_s=0.1$. This yields an astroid PSF that is relatively small, convenient for visual presentation.

Once the telescope is within a distance comparable to the size of the astroid caustic (in this case, within 10 meters), the secondary image begins to widen into an arc. Even closer to the optical axis, the arc splits into three distinct spots of light. As the telescope settles on the optical axis, these spots migrate to their final positions on the circumference of the Einstein ring, resulting in a fully formed Einstein cross. (This simulation assumed that the telescope approaches from one of the principal directions of the astroid caustic, i.e., one of the cusps. To see what happens when the telescope approaches from a different angle, see, e.g., \cite{Turyshev-Toth:2021-quartic}.)

It is remarkable that all these animation frames are simply surface density plots of the integral expression given by Eq.~(\ref{eq:Binsc}), which accurately describes an axisymmetric gravitational lens dominated by a spherically symmetric gravitational potential in all regions, both near and far the optical axis. We can generate with equal ease images seen through a telescope that is positioned as far as a million kilometers or more from the optical axis or a telescope that is at the optical axis or its immediate vicinity.

\subsection{Viewing an extended object}

\begin{figure}
\includegraphics{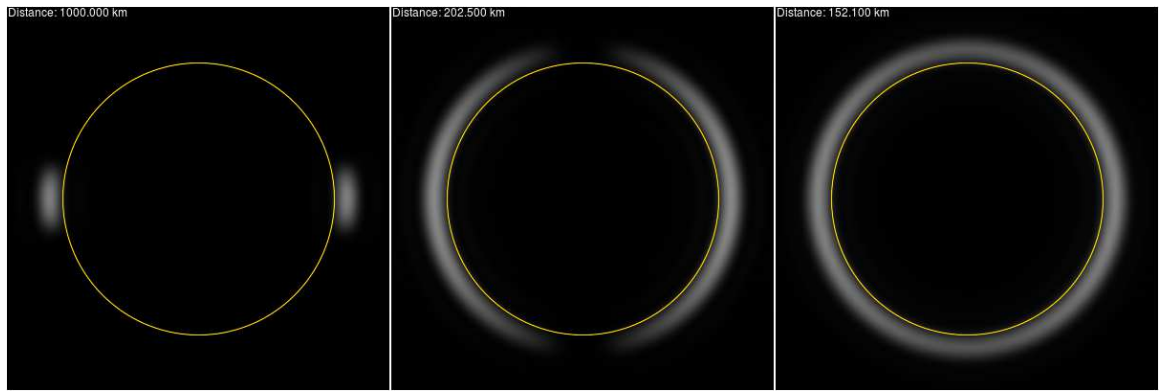}
\caption{\label{fig:hoststar-new}View of a distant star by a telescope approaching the SGL optical axis, at distances of 1,000~km, $\sim$200~km and $\sim$150~km. The geometric projection of the start to the image plane would yield a disk with a radius of 200~km. As the imaging telescope approaches this distance, a full Einstein-ring forms; subsequently, the ring brightens and becomes uniform as the telescope settles on the optical axis.}
\end{figure}

The PSF of a lens represents its impulse response: the image that forms when the light source is a point source. An extended object can, of course, be considered as a collection of point sources. The most straightforward method (though computationally inefficient) of convolving an extended source with the PSF of the lens is by dividing the source into point sources and iterating through them.

To demonstrate this, we considered an extended source in the form of a uniformly illuminated disk, which could represent a host star. We chose a disk that would be geometrically projected to an image with a 200~km radius in the image plane. With the image plane at 650~AU, this would correspond to a Sun-sized star at $\sim$36 light-years.

For computational efficiency, we modeled the extended source using a simple adaptive mesh implementation, refining the resolution for regions that are projected close to the telescope's location in the image plane. This approach was sufficient to create a series of animation frames\footnote{For the full animation, see \url{https://www.vttoth.com/CMS/physics-notes/360}.}, several of which are shown in Fig.~\ref{fig:hoststar-new}.

When the telescope looking at such an extended object is far from the optical axis, the telescopic image appears similar to that produced by a point source (see Fig.~\ref{fig:appr-pre}). However, when the telescope begins to approach the projected image area corresponding to the extended source, a very different picture emerges. Instead of developing into an Einstein cross, the view of the telescope shows a fully formed Einstein ring. We may think of this Einstein ring as a collection of a large number of overlapping Einstein crosses at various orientations, corresponding to the point sources constituting the extended source. Thus, instead of being dominated by light from a single point-like region in the source, the Einstein ring now contains a mix of light from many different regions of the extended source.

\section{Discussion and Conclusions}
\label{sec:end}

We studied the optical properties of an extended axisymmetric gravitational lens. The gravitational potential for such a lens can be described using an infinite series of zonal harmonics. We extended the description of the SGL optical properties from the strong interference region to all lensing regimes. The new results can now also describe lensing in the weak interference region and that in the geometric optics region.

The expressions that we obtained can be used to describe the light field that is created by the SGL in its focal region. It can also be convolved with a representation of an optical telescope (modeled as a thin lens telescope) to show the view seen by such a telescope. The results are ``actionable'' in the sense that they are reduced to a single integral expression that can be evaluated in many cases using direct numerical methods.

Moreover, when used in conjunction with our earlier work \cite{Turyshev-Toth:2021-caustics} in which we obtained a closed form expression of the SGL PSF monopole and quadrupole contribution (ignoring higher-order zonal harmonics that contribute little) the new formalism allows us to compute the light field of the SGL or the view seen by a model telescope without resorting to numerical integration, and thus not hindered by the properties of rapidly oscillating integrals.

We put these results to use, in particular, by creating a series of multiframe animations that show the view of a pont source through a telescope that is approaching the SGL optical axis from afar. The strength of our formalism is powerfully demonstrated when we consider that the same expression can model the (essentially unamplified) view of a distant object when the telescope is still far from the SGL optical axis; the emergence of a secondary image from behind the solar limb; and the eventual widening of these images into arcs and their transition to form an Einstein cross around the Sun. We can also simulate light from extended objects, showing how, even in the presence of multipole moments, such objects still form an Einstein ring around the Sun.

Finally, we note that although our focus remains the SGL that can be represented elegantly using zonal harmonics, our approach can be readily extended to other gravitational lenses that can be represented using symmetric trace-free (STF) tensors \cite{Turyshev-Toth:2021-multipoles}. The resulting formalism covers every gravitational lens that can be described by small deviations from the spherically symmetric gravitational field of a mass monopole. Our approach, therefore, is the most comprehensive wave-theoretical treatment of gravitational lensing in a weak gravitational field to date.

Concluding, we emphasize that the analytical expressions derived in this paper are presented in terms of physically observable quantities and, as such, they are directly suitable for realistic data analysis. To that extent, we can use them to process, e.g., time series brightness data available from the OGLE\footnote{\url{https://en.wikipedia.org/wiki/Optical_Gravitational_Lensing_Experiment}} and MACHO\footnote{\url{https://en.wikipedia.org/wiki/MACHO_Project}}
projects, the upcoming Roman Space Telescope\footnote{\url{https://roman.gsfc.nasa.gov/}}, or other microlensing projects that may benefit from the improved modeling.  In addition, the results presented in this paper offer a solution for establishing a local reference frame that can be used to achieve the required navigational precision for future missions to the SGL's focal region for high-resolution exoplanet imaging \cite{Turyshev-etal:2020-PhaseII}. The corresponding efforts are under way; results, when available, will be published elsewhere.

\begin{acknowledgments}
This work in part was performed at the Jet Propulsion Laboratory, California Institute of Technology, under a contract with the National Aeronautics and Space Administration.
VTT acknowledges the generous support of Plamen Vasilev and other Patreon patrons.
%\textcopyright 2021. All rights reserved.
\end{acknowledgments}

%\bibliography{SGL-diffract}

\appendix
\section{Considering limiting cases}
\label{sec:lim-cases}

Given the complex structure of the results obtained, it is natural to consider limiting cases of the results obtained in this paper for $B\big(\vec x) $ and $ {\cal B} \big(\vec x,\vec x_i\big) $ that are given by (\ref{eq:DB-sol-in-cc+}) and (\ref{eq:Binsc}), correspondingly. The obvious such cases are those for very small deviations $\rho$ from the optical axis, namely $\rho/r\equiv \theta\ll \sqrt{2r_g/r}$, those for very large deviations $\rho/r\equiv \theta\gg \sqrt{2r_g/r}$ and those in between. Below, we will consider each of these cases and will establish correspondence of our results to those studied previously.

\subsection{Small deviations from the optical axis}
\label{sec:small-dev}

We begin with the case of when the deviations from the optical axis are small. In the case when $\rho/r\equiv \theta\ll \sqrt{2r_g/r}$, expression for $a_{\tt }(\vec x,\vec n_\xi) $ given by (\ref{eq:q_insc}) behaves as
{}
  \begin{eqnarray}
\lim_{\theta\rightarrow0}a_{\tt }(\vec x,\vec n_\xi) &=& \lim_{\theta\rightarrow0}\sqrt{\pi k\tilde r}\Bigg[\frac{\Big(\sqrt{\big({\textstyle\frac{1}{2}} ({\vec n}_\xi\cdot{\vec x})/r \big)^2+\frac{2r_g}{ \tilde r}}+{\textstyle\frac{1}{2}}  ({\vec n}_\xi\cdot{\vec x})/r \Big)^3}{\sqrt{\big({\textstyle\frac{1}{2}}  ({\vec n}_\xi\cdot{\vec x})/r \big)^2+\frac{2r_g}{ \tilde r}}}\Bigg]^{1/2}=
\sqrt{2\pi k r_g}+{\cal O}\Big(\frac{\rho}{\sqrt{2r_g\tilde r}}\Big).~~~~
    \label{eq:q_insc-lim}
    \end{eqnarray}

Similarly, we  determine the behavior of the phase shift $\delta\varphi$ from (\ref{eq:Ain-d_ph}):
  \begin{eqnarray}
\lim_{\theta\rightarrow0}    \delta\varphi_{\tt }
(\vec x,\vec n_\xi) &=&-k\lim_{\theta\rightarrow0} \bigg\{{\textstyle\frac{1}{2}}  ({\vec n}_\xi\cdot{\vec x}) \Big( \sqrt{\big({\textstyle\frac{1}{2}}  ({\vec n}_\xi\cdot{\vec x})/r \big)^2+\frac{2r_g}{\tilde r}}+{\textstyle\frac{1}{2}}  ({\vec n}_\xi\cdot{\vec x})/r\Big)+\nonumber\\
&&\hskip -80pt +\,\,2r_g\ln \Big(\sqrt{\big({\textstyle\frac{1}{2}}  ({\vec n}_\xi\cdot{\vec x})/r\big)^2+\frac{2r_g}{\tilde r}}+ {\textstyle\frac{1}{2}}  ({\vec n}_\xi\cdot{\vec x})/r\Big)+
2r_g\sum_{n=2}^\infty  \frac{J_n}{n}\frac{R^2_\odot \sin^n\beta_s\cos[n(\phi_\xi-\phi_s)]}{\Big(\sqrt{\big({\textstyle\frac{1}{2}} ({\vec n}_\xi\cdot{\vec x})\big)^2+2r_g \tilde r}+ {\textstyle\frac{1}{2}} ({\vec n}_\xi\cdot{\vec x})\Big)^n}\bigg\}=\nonumber\\
&&\hskip -80pt =\,
-k\Big\{\sqrt{\frac{2r_g}{\tilde r}}\Big( \rho\cos(\phi_\xi-\phi)+
\sqrt{2r_g\tilde r}\sum_{n=2}^\infty  \frac{J_n}{n}\Big(\frac{R}{\sqrt{2r_g\tilde r}} \Big)^n \sin^n\beta_s\cos[n(\phi_\xi-\phi_s)]+
r_g\ln \frac{2r_g}{\tilde r}+
{\cal O}\Big(\frac{\rho}{\sqrt{2r_g\tilde r}}\Big)\Big\}.
    \label{eq:Ain-d_ph-lim}
\end{eqnarray}

As a result, expressions from the complext amplitude of the EM wave, $ B \big(\vec x\big) $ from (\ref{eq:DB-sol-in-cc+}), and  its Fourier-transform, $ {\cal B} \big(\vec x,\vec x_i\big)$ from (\ref{eq:Binsc}), take familiar forms:
{}
\begin{eqnarray}
 B \big(\vec x\big)
&=&e^{-ikr_g\ln {2r_g}/{\tilde r}}
\sqrt{2\pi k r_g} \,
\times\nonumber\\
&\times& \frac{1}{2\pi}\int_0^{2\pi} d\phi_\xi
\exp\Big[-ik\sqrt{\frac{2r_g}{\tilde r}}\Big( \rho\cos(\phi_\xi-\phi)+
\sqrt{2r_g\tilde r}\sum_{n=2}^\infty  \frac{J_n}{n}\Big(\frac{R}{\sqrt{2r_g\tilde r}} \Big)^n \sin^n\beta_s\cos[n(\phi_\xi-\phi_s)]\Big)\Big],~~~~~
  \label{eq:DB-sol-in-ccLL}
\end{eqnarray}
which was originally obtained in \cite{Turyshev-Toth:2021-multipoles,Turyshev-Toth:2021-caustics} and
{}
\begin{eqnarray}
 {\cal B} \big(\vec x,\vec x_i\big) &=&e^{-ikr_g\ln {2r_g}/{\tilde r}}
\sqrt{2\pi k r_g} \,\frac{1}{2\pi}\int_0^{2\pi} d\phi_\xi \,
  \Big(\frac{
2J_1(u({\vec x},\vec x_i,\vec n_\xi)\frac{1}{2}d)}{u({\vec x},\vec x_i,\vec n_\xi) \frac{1}{2}d}\Big)\times\nonumber\\
&&\hskip -20pt
\times\,
\exp\Big[-ik\sqrt{\frac{2r_g}{\tilde r}}\Big( \rho\cos(\phi_\xi-\phi)+
\sqrt{2r_g\tilde r}\sum_{n=2}^\infty  \frac{J_n}{n}\Big(\frac{R}{\sqrt{2r_g\tilde r}} \Big)^n \sin^n\beta_s\cos[n(\phi_\xi-\phi_s)]\Big)\Big],~~~~~
  \label{eq:BinscER}
\end{eqnarray}
which was obtained in \cite{Turyshev-Toth:2021-imaging,Turyshev-Toth:2021-quartic}.
Therefore, the expressions that we obtained for the complex amplitude of the EM field, $B \big(\vec x\big)$, and its Fourier-transform corresponding to the EM field on the sensor behind a thin lens, $ {\cal B} \big(\vec x,\vec x_i\big)$, are identical to those that we derived earlier \cite{Turyshev-Toth:2021-multipoles,Turyshev-Toth:2021-imaging}.

\subsubsection{Behavior outside the cusps}

 Next, we examine  behavior of $\delta\varphi$ from (\ref{eq:Ain-d_ph}) in the region just outside the caustics. We realize that the term with the multipoles in this region will have a negligible value compared to the leading term in that expression and, thus, it may be omitted, yielding
  \begin{eqnarray}
\delta\varphi_0(\vec x,\vec n_\xi) &=&
-k\bigg\{ {\textstyle\frac{1}{2}}  \tilde r \theta \cos(\phi_\xi-\phi)\Big( \sqrt{\big({\textstyle\frac{1}{2}}  \theta \cos(\phi_\xi-\phi)\big)^2+\frac{2r_g}{\tilde r}}+{\textstyle\frac{1}{2}}  \theta \cos(\phi_\xi-\phi)\Big)+\nonumber\\
&&\hskip 40pt +\,\,2r_g\ln \Big(\sqrt{\big({\textstyle\frac{1}{2}}  \theta \cos(\phi_\xi-\phi)\big)^2+\frac{2r_g}{\tilde r}}+ {\textstyle\frac{1}{2}}  \theta \cos(\phi_\xi-\phi)\Big)+{\cal O}\Big(J_n\Big)
\bigg\}.
    \label{eq:Ain-d_ph-LL}
\end{eqnarray}
This is the phase of the EM wave in the case of a monopole gravitational field, familiar to us from \cite{Turyshev-Toth:2019-extend}.

As a result, expressions from the complex amplitude of the EM wave $ B \big(\vec x\big) $ from (\ref{eq:DB-sol-in-cc+}) and  its Fourier-transform, $ {\cal B} \big(\vec x,\vec x_i\big)$, from (\ref{eq:Binsc}) take the form:
{}
\begin{eqnarray}
 B \big(\vec x\big)
&=& \frac{1}{2\pi}\int_0^{2\pi} d\phi_\xi \,
a_{\tt }(\vec x,\vec n_\xi)
\exp\Big[-ik\Big\{{\textstyle\frac{1}{2}}  \tilde r \theta \cos(\phi_\xi-\phi)\Big( \sqrt{\Big({\textstyle\frac{1}{2}}  \theta \cos(\phi_\xi-\phi)\Big)^2+\frac{2r_g}{\tilde r}}+{\textstyle\frac{1}{2}}  \theta \cos(\phi_\xi-\phi)\Big)+\nonumber\\
&&\hskip 140pt +\,\,2r_g\ln \Big(\sqrt{\big({\textstyle\frac{1}{2}}  \theta \cos(\phi_\xi-\phi)\big)^2+\frac{2r_g}{\tilde r}}+ {\textstyle\frac{1}{2}}  \theta \cos(\phi_\xi-\phi)\Big)\Big\}\Big],~~~~~
  \label{eq:DB-sol-in-ccLLGG}
\end{eqnarray}
and
{}
\begin{eqnarray}
 {\cal B} \big(\vec x,\vec x_i\big) &=& \frac{1}{2\pi}\int_0^{2\pi} d\phi_\xi  \, a_{\tt }(\vec x,\vec n_\xi)
  \Big(\frac{
2J_1(u({\vec x},\vec x_i,\vec n_\xi)\frac{1}{2}d)}{u({\vec x},\vec x_i,\vec n_\xi) \frac{1}{2}d}\Big)\times\nonumber\\
&\times&
\exp\Big[-ik\Big\{{\textstyle\frac{1}{2}}  \tilde r \theta \cos(\phi_\xi-\phi)\Big( \sqrt{\big({\textstyle\frac{1}{2}}  \theta \cos(\phi_\xi-\phi)\big)^2+\frac{2r_g}{\tilde r}}+{\textstyle\frac{1}{2}}  \theta \cos(\phi_\xi-\phi)\Big)+\nonumber\\
&&\hskip 50pt +\,\,2r_g\ln \Big(\sqrt{\big({\textstyle\frac{1}{2}}  \theta \cos(\phi_\xi-\phi)\big)^2+\frac{2r_g}{\tilde r}}+ {\textstyle\frac{1}{2}}  \theta \cos(\phi_\xi-\phi)\Big)\Big\}\Big].~~~~~
  \label{eq:BinscER_GG}
\end{eqnarray}

In the region outside the caustic, we can take the two integrals (\ref{eq:DB-sol-in-ccLLGG}) and (\ref{eq:BinscER_GG}) using the method of stationary phase. In both of these expressions, we are dealing with the same phase $\delta\varphi_0(\vec x,\vec n_\xi)$ given by (\ref{eq:Ain-d_ph-LL}). The phase is stationary when $d\delta\varphi_0(\vec x,\vec n_\xi)/\phi_\xi=0$. This condition yields two solutions $\phi_\xi-\phi=0$ and $\phi_\xi-\phi=\pi$. Computing  $d^2\delta\varphi_0(\vec x,\vec n_\xi)/d\phi_\xi^2$ for both cases, we obtain
{}
\begin{eqnarray}
\frac{d^2\delta\varphi_0(\vec x,\vec n_\xi)}{d\phi_\xi^2}\Big|_{\phi_\xi-\phi=0} &=&
k\tilde r \theta\Big( \sqrt{\big({\textstyle\frac{1}{2}} \theta \big)^2+\frac{2r_g}{ \tilde r}}+{\textstyle\frac{1}{2}}  \theta \Big) +{\cal O}(r_g^2),
  \label{eq:phi-dd1}\\
\frac{d^2\delta\varphi_0(\vec x,\vec n_\xi)}{d\phi_\xi^2}\Big|_{\phi_\xi-\phi=\pi} &=&
-k\tilde r \theta\Big( \sqrt{\big({\textstyle\frac{1}{2}} \theta \big)^2+\frac{2r_g}{ \tilde r}}-{\textstyle\frac{1}{2}}  \theta\Big) +
{\cal O}(r_g^2).
  \label{eq:phi-dd2}
\end{eqnarray}

Now we consider behavior of the expression for $a_{\tt }(\vec x,\vec n_\xi) $ given by (\ref{eq:q_insc})
{}
  \begin{eqnarray}
a_{\tt }(\vec x,\vec n_\xi)\big|_{\phi_\xi-\phi=0} &=& \sqrt{\pi k\tilde r}\Bigg[\frac{\Big(\sqrt{\big({\textstyle\frac{1}{2}} \theta  \big)^2+\frac{2r_g}{ \tilde r}}+{\textstyle\frac{1}{2}}  \theta \Big)^3}{\sqrt{\big({\textstyle\frac{1}{2}}  \theta  \big)^2+\frac{2r_g}{ \tilde r}}}\Bigg]^{1/2},
    \label{eq:q_insc-lim_L0}\\
    a_{\tt }(\vec x,\vec n_\xi)\big|_{\phi_\xi-\phi=\pi} &=& \sqrt{\pi k\tilde r}\Bigg[\frac{\Big(\sqrt{\big({\textstyle\frac{1}{2}} \theta \big)^2+\frac{2r_g}{ \tilde r}}-{\textstyle\frac{1}{2}}  \theta \Big)^3}{\sqrt{\big({\textstyle\frac{1}{2}}  \theta \big)^2+\frac{2r_g}{ \tilde r}}}\Bigg]^{1/2}.
    \label{eq:q_insc-lim_L02}
    \end{eqnarray}

These expressions may be evaluated in two different regions, namely
\begin{inparaenum}[1)]
\item the region just outside the cusp, but still within the strong interference region,  and
\item the region at a significant distance from the optical axis in the regions of weak interference and that of geometric optics.
\end{inparaenum} These expressions are identical to those obtained in \cite{Turyshev-Toth:2019-image,Turyshev-Toth:2020-im-extend}.

\subsubsection{Larger deviations from the optical axis, but outside the caustic}

Consider studying the region at larger distances outside the caustic, but still within the strong interference region. In this case, $\rho/r\equiv \theta\ll \sqrt{2r_g/r}$, yielding an appropriate small parameter $\theta/\sqrt{2r_g/r}$. We will use this parameter to simplify the results obtained above.
Under these conditions, expressions (\ref{eq:q_insc-lim_L0})  and (\ref{eq:q_insc-lim_L02}) behave as
{}
  \begin{eqnarray}
\lim_{\theta/\sqrt{2r_g/r}\rightarrow0}\Big(a_{\tt }(\vec x,\vec n_\xi)\big|_{\phi_\xi-\phi=0}\Big) &=& \lim_{\theta/\sqrt{2r_g/r}\rightarrow0}\sqrt{\pi k\tilde r}\Bigg[\frac{\Big(\sqrt{\big({\textstyle\frac{1}{2}} \theta \big)^2+\frac{2r_g}{ \tilde r}}+{\textstyle\frac{1}{2}}  \theta \Big)^3}{\sqrt{\big({\textstyle\frac{1}{2}}  \theta  \big)^2+\frac{2r_g}{ \tilde r}}}\Bigg]^{1/2}=\nonumber\\
&=&\sqrt{2\pi k r_g}\Big\{1+{\textstyle\frac{3}{4}}\frac{\theta}{\sqrt{2r_g/ \tilde r}} +{\textstyle\frac{7}{32}} \Big(\frac{\theta}{\sqrt{2r_g/\tilde r}}\Big)^2+{\cal O}(r_g^2,\theta^3)\Big\},
    \label{eq:q_insc-lim_L10}\\
\lim_{\theta/\sqrt{2r_g/r}\rightarrow0}\Big(a_{\tt }(\vec x,\vec n_\xi)\big|_{\phi_\xi-\phi_0=\pi}\Big) &=& \lim_{\theta/\sqrt{2r_g/r}\rightarrow0}\sqrt{\pi k\tilde r}\Bigg[\frac{\Big(\sqrt{\big({\textstyle\frac{1}{2}} \theta \big)^2+\frac{2r_g}{ \tilde r}}-{\textstyle\frac{1}{2}}  \theta \Big)^3}{\sqrt{\big({\textstyle\frac{1}{2}}  \theta  \big)^2+\frac{2r_g}{ \tilde r}}}\Bigg]^{1/2}=\nonumber\\
&=&\sqrt{2\pi k r_g}\Big\{1-{\textstyle\frac{3}{4}}\frac{\theta}{\sqrt{2r_g/ \tilde r}} +{\textstyle\frac{7}{32}} \Big(\frac{\theta}{\sqrt{2r_g/ \tilde r}}\Big)^2+{\cal O}(r_g^2,\theta^3)\Big\}.
    \label{eq:q_insc-lim_L20}
    \end{eqnarray}

The second derivative of the phase is computed from (\ref{eq:phi-dd1}) and (\ref{eq:phi-dd2}) as
{}
\begin{eqnarray}
\lim_{\theta/\sqrt{2r_g/r}\rightarrow0}\Big(\frac{d^2\delta\varphi_0(\vec x,\vec n_\xi)}{d\phi_\xi^2}\Big|_{\phi_\xi-\phi=0} \Big)&=&
\lim_{\theta/\sqrt{2r_g/r}\rightarrow0}\Big\{k\tilde r \theta\Big( \sqrt{\big({\textstyle\frac{1}{2}} \theta \big)^2+\frac{2r_g}{ \tilde r}}+{\textstyle\frac{1}{2}}  \theta \Big) +{\cal O}(r_g^2)\Big)\Big\}=\nonumber\\
&=&k\sqrt{2r_g\tilde r}\theta\Big\{1+{\textstyle\frac{1}{2}}\frac{\theta}{\sqrt{2r_g/ \tilde r}} +{\textstyle\frac{1}{8}} \Big(\frac{\theta}{\sqrt{2r_g/ \tilde r}}\Big)^2+{\cal O}(r_g^2,\theta^3)\Big\},
  \label{eq:phi-dd12*}\\
\lim_{\theta/\sqrt{2r_g/r}\rightarrow0}\Big(\frac{d^2\delta\varphi_0(\vec x,\vec n_\xi)}{d\phi_\xi^2}\Big|_{\phi_\xi-\phi=\pi} \Big)&=&
\lim_{\theta/\sqrt{2r_g/r}\rightarrow0}\Big\{-k\tilde r \theta\Big( \sqrt{\big({\textstyle\frac{1}{2}} \theta\big)^2+\frac{2r_g}{ \tilde r}}-{\textstyle\frac{1}{2}}  \theta \Big) +
{\cal O}(r_g^2)\Big)\Big\}=\nonumber\\
&=&-k\sqrt{2r_g\tilde r}\theta\Big\{1-{\textstyle\frac{1}{2}}\frac{\theta}{\sqrt{2r_g/ \tilde r}} +{\textstyle\frac{1}{8}} \Big(\frac{\theta}{\sqrt{2r_g/ \tilde r}}\Big)^2+{\cal O}(r_g^2,\theta^3)\Big\}.
  \label{eq:phi-dd22*}
\end{eqnarray}

This allows us to compute
{}
\begin{eqnarray}
\frac{1}{2\pi }a_{\tt }(\vec x,\vec n_\xi)\sqrt{\frac{2\pi}{|\delta\varphi_0''|}}\Big|_{\phi_\xi-\phi=0}&=&
\bigg(\frac{\sqrt{2r_g\tilde r}}{2\tilde r \theta}\bigg)^\frac{1}{2}\Big\{1+{\textstyle\frac{1}{2}}\frac{\tilde r \theta}{\sqrt{2r_g\tilde r}} +{\textstyle\frac{1}{16}} \Big(\frac{\tilde r\theta}{\sqrt{2r_g \tilde r}}\Big)^2+{\cal O}(r_g^2,\theta^3)\Big\},
  \label{eq:phi-aa10}\\
\frac{1}{2\pi }a_{\tt }(\vec x,\vec n_\xi)\sqrt{\frac{2\pi}{|\delta\varphi_0''|}}\Big|_{\phi_\xi-\phi=\pi} &=&
\bigg(\frac{\sqrt{2r_g\tilde r}}{2\tilde r \theta}\bigg)^\frac{1}{2}\Big\{1-{\textstyle\frac{1}{2}}\frac{\tilde r \theta}{\sqrt{2r_g\tilde r}} +{\textstyle\frac{1}{16}} \Big(\frac{\tilde r\theta}{\sqrt{2r_g \tilde r}}\Big)^2+{\cal O}(r_g^2,\theta^3)\Big\}.
  \label{eq:phi-aa20}
\end{eqnarray}

Finally, the phase $\delta\varphi_0(\vec x) $ from (\ref{eq:Ain-d_ph-LL})  for the two solutions  takes the form
{}
  \begin{eqnarray}
\delta\varphi_0(\vec x)\big|_{\phi_\xi-\phi=0} &=&
-2kr_g\Big\{\ln\sqrt{\frac{2r_g}{\tilde r}}+\frac{\theta}{\sqrt{2r_g/\tilde r}} +{\textstyle\frac{1}{4}} \Big(\frac{\theta}{\sqrt{2r_g/\tilde r}}\Big)^2+{\cal O}(r_g^2,J_n,\theta^3)\Big\} \equiv
\delta\hat\varphi_{\tt in}(\vec x),
    \label{eq:Ain-d_ph-LL+10}\\[4pt]
\delta\varphi_0(\vec x)\big|_{\phi_\xi-\phi=\pi} &=&
-2kr_g\Big\{\ln\sqrt{\frac{2r_g}{\tilde r}}-\frac{\theta}{\sqrt{2r_g/\tilde r}} +{\textstyle\frac{1}{4}} \Big(\frac{\theta}{\sqrt{2r_g/\tilde r}}\Big)^2+{\cal O}(r_g^2,J_n,\theta^3)\Big\}\equiv\delta\hat\varphi_{\tt sc}(\vec x).
    \label{eq:Ain-d_ph-LL+20}
\end{eqnarray}

Therefore, expressions for the complex amplitude of the EM wave, $B \big(\vec x\big) $ from (\ref{eq:DB-sol-in-ccLLGG}), and  its Fourier-transform, $ {\cal B} \big(\vec x,\vec x_i\big)$ from (\ref{eq:BinscER_GG}), take the form:
{}
\begin{eqnarray}
 B \big(\vec x\big)
&=&
\frac{1}{2}
\bigg(\sqrt{\frac{2r_g}{\tilde r}}\frac{1}{{\textstyle\frac{1}{2}}\theta}\bigg)^\frac{1}{2}\Big\{1+\frac{{\textstyle\frac{1}{2}}\theta}{\sqrt{2r_g/\tilde r}} +{\textstyle\frac{1}{4}} \Big(\frac{{\textstyle\frac{1}{2}}\theta}{\sqrt{2r_g/\tilde r}}\Big)^2\Big\}e^{i\big(\delta\hat\varphi_{\tt in}(\vec x)+{\textstyle\frac{\pi}{4}}\big)}+\nonumber\\
&&\hskip 60pt
+\, \frac{1}{2}\bigg(\sqrt{\frac{2r_g}{\tilde r}}\frac{1}{{\textstyle\frac{1}{2}}\theta}\bigg)^\frac{1}{2}\Big\{1-\frac{{\textstyle\frac{1}{2}}\theta}{\sqrt{2r_g/\tilde r}} +{\textstyle\frac{1}{4}} \Big(\frac{{\textstyle\frac{1}{2}}\theta}{\sqrt{2r_g/\tilde r}}\Big)^2\Big\}e^{i\big(\delta\hat\varphi_{\tt sc}(\vec x)-{\textstyle\frac{\pi}{4}}\big)}+{\cal O}(r_g^2),
  \label{eq:DB-sol-in-ccLLGG20}
\end{eqnarray}
and
{}
\begin{eqnarray}
 {\cal B} \big(\vec x,\vec x_i\big) &=&  \frac{1}{2}
\bigg(\sqrt{\frac{2r_g}{\tilde r}}\frac{1}{{\textstyle\frac{1}{2}}\theta}\bigg)^\frac{1}{2}\Big\{1+\frac{{\textstyle\frac{1}{2}}\theta}{\sqrt{2r_g/\tilde r}} +{\textstyle\frac{1}{4}} \Big(\frac{{\textstyle\frac{1}{2}}\theta}{\sqrt{2r_g/\tilde r}}\Big)^2\Big\}
 \Big(\frac{
2J_1(\hat u_{\tt in}({\vec x},\vec x_i)\frac{1}{2}d)}{\hat u_{\tt in}({\vec x},\vec x_i) \frac{1}{2}d}\Big)
e^{i\big(\delta\hat\varphi_{\tt in}(\vec x)+{\textstyle\frac{\pi}{4}}\big)}+\nonumber\\
&&\hskip 0pt
+\,\frac{1}{2}\bigg(\sqrt{\frac{2r_g}{\tilde r}}\frac{1}{{\textstyle\frac{1}{2}}\theta}\bigg)^\frac{1}{2}\Big\{1-\frac{{\textstyle\frac{1}{2}}\theta}{\sqrt{2r_g/\tilde r}} +{\textstyle\frac{1}{4}} \Big(\frac{{\textstyle\frac{1}{2}}\theta}{\sqrt{2r_g/\tilde r}}\Big)^2\Big\}\Big(\frac{
2J_1(\hat u_{\tt sc}({\vec x},\vec x_i)\frac{1}{2}d)}{\hat u_{\tt sc}({\vec x},\vec x_i) \frac{1}{2}d}\Big)e^{i\big(\delta\hat\varphi_{\tt sc}(\vec x)-{\textstyle\frac{\pi}{4}}\big)}+{\cal O}(r_g^2),  ~~~
\label{eq:BinscER_GG20}
\end{eqnarray}
where phases $\delta\hat\varphi_{\tt in/sc}$  are from (\ref{eq:Ain-d_ph-LL+10})--(\ref{eq:Ain-d_ph-LL+20}) and spatial frequencies $\hat u_{\tt in}({\vec x}_i,\vec x)$ and $\hat u_{\tt sc}({\vec x},\vec x_i)$  from (\ref{eq:vpms}) are given as
{}
\begin{eqnarray}
\hat u_{\tt in/sc}({\vec x},\vec x_i)
&=&\sqrt{\hat \nu_{\tt in/sc}^2\pm2\hat \nu_{\tt in/sc}\eta_i\cos\big(\phi-\phi_i\big)+\eta_i^2},
\label{eq:vpms10}
\end{eqnarray}
and frequency $\hat\nu_{\tt in/sc}(\vec x)$ from (\ref{eq:betapm}) has the from
 {}
 \begin{eqnarray}
\hat\nu_{\tt in/sc}(\vec x)&=&k\sqrt{\frac{2r_g}{\tilde r}}\Big(1\pm  \frac{{\textstyle\frac{1}{2}}\theta}{\sqrt{2r_g/\tilde r}} +{\textstyle\frac{1}{2}} \Big(\frac{{\textstyle\frac{1}{2}}\theta}{\sqrt{2r_g/\tilde r}}\Big)^2+{\cal O}(r_g^2,\theta^3)\Big),
  \label{eq:betapm20}
  \end{eqnarray}
where $'+'$ and $'-'$ signes are for incident $'{\tt in}'$ and scattered $'{\tt sc}'$  waves, correspondingly, and also $\theta=\rho/\tilde r$.  Clearly, these expressions are identical to those obtained in \cite{Turyshev-Toth:2019-extend,Turyshev-Toth:2019-image,Turyshev-Toth:2020-im-extend}.

\subsection{Large deviations from the optical axis}

We now consider the region at a significant distance from the optical axis in the regions of weak interference and that of geometric optics.
In the case $\rho/r\equiv \theta\gg \sqrt{2r_g/r}$, expression (\ref{eq:q_insc-lim_L0})  behaves as below:
{}
  \begin{eqnarray}
\lim_{2r_g/r\theta^2\rightarrow0}a_{\tt }(\vec x,\vec n_\xi)\big|_{\phi_\xi-\phi=0} &=& \lim_{2r_g/r\theta^2\rightarrow0}\sqrt{\pi k\tilde r}\Bigg[\frac{\Big(\sqrt{\big({\textstyle\frac{1}{2}} \theta \big)^2+\frac{2r_g}{ \tilde r}}+{\textstyle\frac{1}{2}}  \theta \Big)^3}{\sqrt{\big({\textstyle\frac{1}{2}}  \theta  \big)^2+\frac{2r_g}{ \tilde r}}}\Bigg]^{1/2}=
\sqrt{2\pi k \tilde r}\Big\{\theta +\frac{r_g}{\tilde r\theta}+{\cal O}(r_g^2)\Big\},
    \label{eq:q_insc-lim_L1}\\
\lim_{2r_g/r\theta^2\rightarrow0}a_{\tt }(\vec x,\vec n_\xi)\big|_{\phi_\xi-\phi=\pi} &=& \lim_{2r_g/r\theta^2\rightarrow0}\sqrt{\pi k\tilde r}\Bigg[\frac{\Big(\sqrt{\big({\textstyle\frac{1}{2}} \theta \big)^2+\frac{2r_g}{ \tilde r}}-{\textstyle\frac{1}{2}}  \theta \Big)^3}{\sqrt{\big({\textstyle\frac{1}{2}}  \theta  \big)^2+\frac{2r_g}{ \tilde r}}}\Bigg]^{1/2}=
\sqrt{\pi k \tilde r}\Big\{\Big(\frac{r_g}{\tilde r}\Big)^{3/2}\frac{4}{\theta^2}+{\cal O}(r_g^{5/2})\Big\}.~~~~
    \label{eq:q_insc-lim_L2}
    \end{eqnarray}

This allows us to compute
{}
\begin{eqnarray}
\frac{1}{2\pi }a_{\tt }(\vec x,\vec n_\xi)\sqrt{\frac{2\pi}{|\delta\varphi_0''|}}\Big|_{\phi_\xi-\phi=0} &=&\frac{1}{2\pi }\sqrt{2\pi k \tilde r}\theta\Big(1 +\frac{r_g}{\tilde r\theta^2}\Big)\sqrt{\frac{2\pi}{k\tilde r \theta^2\big(1+\frac{2r_g}{\tilde r\theta^2}\big)}}=1+{\cal O}(r_g^2),
  \label{eq:phi-aa1}\\
\frac{1}{2\pi }a_{\tt }(\vec x,\vec n_\xi)\sqrt{\frac{2\pi}{|\delta\varphi_0''|}}\Big|_{\phi_\xi-\phi=\pi} &=&\frac{1}{2\pi }\sqrt{\pi k \tilde r}\Big(\frac{r_g}{\tilde r}\Big)^{3/2}\frac{4}{\theta^2}\sqrt{\frac{2\pi}{2kr_g}}=\frac{r_g}{\tilde r{\textstyle\frac{1}{2}}  \theta^2}\simeq \frac{r_g}{\tilde r(1-\cos\theta)}+{\cal O}(r_g^2).
  \label{eq:phi-aa2}
\end{eqnarray}

Finally, the phase $\delta\varphi_0(\vec x) $ from (\ref{eq:Ain-d_ph-LL})  for the two solutions  takes the form
{}
\begin{eqnarray}
\delta\varphi_0(\vec x)\big|_{\phi_\xi-\phi=0} &=&
-k\Big\{ {\textstyle\frac{1}{2}}  \tilde r \theta \Big( \sqrt{\big({\textstyle\frac{1}{2}}  \theta \big)^2+\frac{2r_g}{\tilde r}}+{\textstyle\frac{1}{2}}  \theta \Big)+
2r_g\ln \Big(\sqrt{\big({\textstyle\frac{1}{2}}  \theta \big)^2+\frac{2r_g}{\tilde r}}+ {\textstyle\frac{1}{2}}  \theta \Big)+{\cal O}\big(J_n\big)
\Big\} \equiv
\delta\varphi_{\tt in}(\vec x),
    \label{eq:Ain-d_ph-LL+1}\\[4pt]
\delta\varphi_0(\vec x)\big|_{\phi_\xi-\phi=\pi} &=&
-k\Big\{ -{\textstyle\frac{1}{2}}  \tilde r \theta \Big( \sqrt{\big({\textstyle\frac{1}{2}}  \theta \big)^2+\frac{2r_g}{\tilde r}}-{\textstyle\frac{1}{2}}  \theta \Big)+
2r_g\ln \Big(\sqrt{\big({\textstyle\frac{1}{2}}  \theta \big)^2+\frac{2r_g}{\tilde r}}- {\textstyle\frac{1}{2}}  \theta \Big)+{\cal O}\big(J_n\big)
\Big\}\equiv
\delta\varphi_{\tt sc}(\vec x).~~~
    \label{eq:Ain-d_ph-LL+2}
\end{eqnarray}

Expressions for $\delta\varphi_{\tt in/sc}$ from (\ref{eq:Ain-d_ph-LL+1})--(\ref{eq:Ain-d_ph-LL+2}) may be further simplified taking into account that in this region $\theta\gg \sqrt{2r_g/r}$. Taking this fact into account, we have
{}
  \begin{eqnarray}
\delta\varphi_{\tt in}(\vec x)&=&-k\Big\{r_g\big(1+2\ln\theta\big)+{\textstyle\frac{1}{2}}\tilde r\theta^2+{\cal O}(r_g^2)\Big\},
    \label{eq:Ain-d_ph-LL+13}\\[4pt]
\delta\varphi_{\tt sc}(\vec x) &=&kr_g\Big\{1-2\ln\frac{2r_g}{\tilde r\theta}
+{\cal O}(r_g^2)\Big\}.
    \label{eq:Ain-d_ph-LL+23}
\end{eqnarray}
After combining these results with the $\Omega(t)$ from (\ref{eq:omega}), we obtain phases of the incident and scattered waves with the same structure as was in (23)--(24) of \cite{Turyshev-Toth:2017}, as expected.

Therefore, expressions from the complex amplitude of the EM wave $ B \big(\vec x\big) $ from (\ref{eq:DB-sol-in-ccLLGG}) and  its Fourier-transform, $ {\cal B} \big(\vec x,\vec x_i\big)$, from (\ref{eq:BinscER_GG}) take the form:
{}
\begin{eqnarray}
 B \big(\vec x\big)
&=&
e^{i\big(\delta\hat\varphi_{\tt in}(\vec x)+{\textstyle\frac{\pi}{4}}\big)}
+ \frac{r_g}{\tilde r(1-\cos\theta)}
e^{i\big(\delta\hat\varphi_{\tt sc}(\vec x)-{\textstyle\frac{\pi}{4}}\big)}+{\cal O}(r_g^2),
  \label{eq:DB-sol-in-ccLLGG2}
\end{eqnarray}
and
{}
\begin{eqnarray}
 {\cal B} \big(\vec x,\vec x_i\big) &=&  \Big(\frac{
2J_1(u_{\tt in}({\vec x},\vec x_i)\frac{1}{2}d)}{u_{\tt in}({\vec x},\vec x_i) \frac{1}{2}d}\Big)
e^{i\big(\delta\hat\varphi_{\tt in}(\vec x)+{\textstyle\frac{\pi}{4}}\big)}
+ \frac{r_g}{\tilde r(1-\cos\theta)} \Big(\frac{
2J_1(u_{\tt sc}({\vec x},\vec x_i)\frac{1}{2}d)}{u_{\tt sc}({\vec x},\vec x_i) \frac{1}{2}d}\Big)e^{i\big(\delta\hat\varphi_{\tt sc}(\vec x)-{\textstyle\frac{\pi}{4}}\big)}+{\cal O}(r_g^2),  ~~~
\label{eq:BinscER_GG2}
\end{eqnarray}
where phases $\delta\varphi_{\tt in/sc}$  are from (\ref{eq:Ain-d_ph-LL+1})--(\ref{eq:Ain-d_ph-LL+2}) and spatial frequencies $u_{\tt in}({\vec x},\vec x_i)$ and $u_{\tt sc}({\vec x},\vec x_i)$  from (\ref{eq:vpms}) are given as
{}
\begin{eqnarray}
u_{\tt in/sc}({\vec x},\vec x_i)
&=&\sqrt{\nu_{\tt in/sc}^2\pm2\nu_{\tt in/sc}\eta_i\cos\big(\phi-\phi_i\big)+\eta_i^2},
\label{eq:vpms1}
\end{eqnarray}
and frequencies $\nu_{\tt in/sc}(\vec x)$ from (\ref{eq:betapm}) have the from
 {}
 \begin{eqnarray}
\nu_{\tt in}(\vec x)&=&k\Big(\theta+\frac{2r_g}{\tilde r \theta}\Big)=k\theta\Big(1+\frac{2r_g}{\tilde r(1-\cos\theta)}+{\cal O}(r_g^2,\theta^4)\Big),\\
\nu_{\tt sc}(\vec x)&=&k\frac{2r_g}{\tilde r \theta}=k\theta\Big(\frac{2r_g}{\tilde r(1-\cos\theta)}+{\cal O}(r_g^2,\theta^4)\Big).
  \label{eq:betapm23}
  \end{eqnarray}

These results are identical to those obtained in \cite{Turyshev-Toth:2017,Turyshev-Toth:2019-extend,Turyshev-Toth:2019-image}.

\subsection{Complete description in the area outsize the cusps}

At this moment, we can give a complete description of the EM field in the region outside the cusp. We have established earlier that the amplitudes given by expressions (\ref{eq:q_insc-lim_L0}) and (\ref{eq:q_insc-lim_L02}) behave as below:
{}
  \begin{eqnarray}
a_{\tt }(\vec x,\vec n_\xi)\big|_{\phi_\xi-\phi=0} &=& \sqrt{\pi k\tilde r}\Bigg[\frac{\Big(\sqrt{\big({\textstyle\frac{1}{2}} \theta \big)^2+\frac{2r_g}{ \tilde r}}+{\textstyle\frac{1}{2}}  \theta \Big)^3}{\sqrt{\big({\textstyle\frac{1}{2}}  \theta  \big)^2+\frac{2r_g}{ \tilde r}}}\Bigg]^{1/2},
    \label{eq:q_insc-lim_L15}\\
a_{\tt }(\vec x,\vec n_\xi)\big|_{\phi_\xi-\phi=\pi} &=&\sqrt{\pi k\tilde r}\Bigg[\frac{\Big(\sqrt{\big({\textstyle\frac{1}{2}} \theta \big)^2+\frac{2r_g}{ \tilde r}}-{\textstyle\frac{1}{2}}  \theta \Big)^3}{\sqrt{\big({\textstyle\frac{1}{2}}  \theta  \big)^2+\frac{2r_g}{ \tilde r}}}\Bigg]^{1/2}.
    \label{eq:q_insc-lim_L25}
    \end{eqnarray}

Together with the appropriately approximated (\ref{eq:phi-dd1}) and (\ref{eq:phi-dd2}), this allows us to compute
{}
\begin{eqnarray}
\frac{1}{2\pi }a_{\tt }(\vec x,\vec n_\xi)\sqrt{\frac{2\pi}{|\delta\varphi_0''|}}\Big|_{\phi_\xi-\phi=0} &=&
\frac{1}{2} \frac{\sqrt{1+\frac{8r_g}{ \tilde r\theta^2}}+1}{\Big(1+\frac{8r_g}{ \tilde r\theta^2}\Big)^\frac{1}{4}  }+{\cal O}(r_g^2),
  \label{eq:phi-aa15}\\
\frac{1}{2\pi }a_{\tt }(\vec x,\vec n_\xi)\sqrt{\frac{2\pi}{|\delta\varphi_0''|}}\Big|_{\phi_\xi-\phi=\pi} &=&
\frac{1}{2} \frac{\sqrt{ 1+\frac{8r_g}{ \tilde r \theta^2}}- 1 }{\Big(1+\frac{8r_g}{ \tilde r\theta^2}\Big)^\frac{1}{4}  }+{\cal O}(r_g^2).
  \label{eq:phi-aa25}
\end{eqnarray}

Finally, the phase $\delta\varphi_0(\vec x) $ from (\ref{eq:Ain-d_ph-LL})  for the two solutions  takes the form
{}
  \begin{eqnarray}
\delta\varphi_0(\vec x)\big|_{\phi_\xi-\phi=0} &=&
-k\Big\{ {\textstyle\frac{1}{2}}  \tilde r \theta \Big( \sqrt{\big({\textstyle\frac{1}{2}}  \theta \big)^2+\frac{2r_g}{\tilde r}}+{\textstyle\frac{1}{2}}  \theta \Big)+
2r_g\ln \Big(\sqrt{\big({\textstyle\frac{1}{2}}  \theta \big)^2+\frac{2r_g}{\tilde r}}+ {\textstyle\frac{1}{2}}  \theta \Big)+{\cal O}\big(J_n\big)
\Big\} \equiv
\delta\varphi_{\tt in}(\vec x),
    \label{eq:Ain-d_ph-LL+15}\\[4pt]
\delta\varphi_0(\vec x)\big|_{\phi_\xi-\phi=\pi} &=&
-k\Big\{ -{\textstyle\frac{1}{2}}  \tilde r \theta \Big( \sqrt{\big({\textstyle\frac{1}{2}}  \theta \big)^2+\frac{2r_g}{\tilde r}}-{\textstyle\frac{1}{2}}  \theta \Big)+
2r_g\ln \Big(\sqrt{\big({\textstyle\frac{1}{2}}  \theta \big)^2+\frac{2r_g}{\tilde r}}- {\textstyle\frac{1}{2}}  \theta \Big)+{\cal O}\big(J_n\big)
\Big\}\equiv
\delta\varphi_{\tt sc}(\vec x).~~~
    \label{eq:Ain-d_ph-LL+25}
\end{eqnarray}

Therefore, expressions from the complex amplitude of the EM wave $ B \big(\vec x\big) $ from (\ref{eq:DB-sol-in-ccLLGG}) and  its Fourier-transform, $ {\cal B} \big(\vec x,\vec x_i\big)$, from (\ref{eq:BinscER_GG}) take the form:
{}
\begin{eqnarray}
 B \big(\vec x\big)
&=&
\frac{1}{2} \frac{\sqrt{1+\frac{8r_g}{ \tilde r\theta^2}}+ 1}{\Big(1+\frac{8r_g}{ \tilde r \theta^2}\Big)^\frac{1}{4}  }e^{i\big(\delta\varphi_{\tt in}(\vec x)+{\textstyle\frac{\pi}{4}}\big)}
+
\frac{1}{2} \frac{\sqrt{ 1+\frac{8r_g}{ \tilde r \theta^2}}- 1 }{\Big(1+\frac{8r_g}{ \tilde r\theta^2}\Big)^\frac{1}{4}  }
e^{i\big(\delta\varphi_{\tt sc}(\vec x)-{\textstyle\frac{\pi}{4}}\big)}+{\cal O}(r_g^2),
  \label{eq:DB-sol-in-ccLLGG25}
\end{eqnarray}
and
\begin{eqnarray}
 {\cal B} \big(\vec x,\vec x_i\big) &=& \frac{1}{2} \frac{\sqrt{1+\frac{8r_g}{ \tilde r\theta^2}}+ 1}{\Big(1+\frac{8r_g}{ \tilde r\theta^2}\Big)^\frac{1}{4} } \Big(\frac{
2J_1(u_{\tt in}({\vec x},\vec x_i)\frac{1}{2}d)}{u_{\tt in}({\vec x},\vec x_i) \frac{1}{2}d}\Big)
 e^{i\big(\delta\varphi_{\tt in}(\vec x)+{\textstyle\frac{\pi}{4}}\big)}+
\nonumber\\&&\hskip 40pt +\,
\frac{1}{2} \frac{\sqrt{ 1+\frac{8r_g}{ \tilde r \theta^2}}- 1 }{\Big(1+\frac{8r_g}{ \tilde r \theta^2}\Big)^{\frac{1}{4}}}\Big(\frac{
2J_1(u_{\tt sc}({\vec x},\vec x_i)\frac{1}{2}d)}{u_{\tt sc}({\vec x},\vec x_i) \frac{1}{2}d}\Big)e^{i\big(\delta\varphi_{\tt sc}(\vec x)-{\textstyle\frac{\pi}{4}}\big)}+{\cal O}(r_g^2),  ~~~~
\label{eq:BinscER_GG25}
\end{eqnarray}
where phases $\delta\varphi_{\tt in/sc}$  are from (\ref{eq:Ain-d_ph-LL+1})--(\ref{eq:Ain-d_ph-LL+2}) and spatial frequencies $u_{\tt in}({\vec x},\vec x_i)$ and $u_{\tt sc}({\vec x},\vec x_i)$  from (\ref{eq:vpms}) are given as
{}
\begin{eqnarray}
u_{\tt in/sc}({\vec x},\vec x_i)
&=&\sqrt{\nu_{\tt in/sc}^2\pm2\nu_{\tt in/sc}\eta_i\cos\big(\phi-\phi_i\big)+\eta_i^2},
\label{eq:vpms15}
\end{eqnarray}
and frequency $\nu_{\tt in/sc}(\vec x)$ from (\ref{eq:betapm}) has the from
 {}
 \begin{eqnarray}
\nu_{\tt in/sc}(\vec x)&=&
k{\textstyle\frac{1}{2}}\Big(\sqrt{\theta^2+\frac{8r_g}{ \tilde r}}\pm  \theta\Big),
  \label{eq:betapm25}
  \end{eqnarray}
where the positive and negative signs are for incident (${\tt in}$) and scattered (${\tt sc}$)  waves, correspondingly. As we mentioned earlier, these results are identical to those obtained in \cite{Turyshev-Toth:2017,Turyshev-Toth:2019-extend,Turyshev-Toth:2019-image}. However, the results reported in this paper allow us to generalize the description of the gravitational lensing phenomena and use the same expression in all the regions of interest, thus providing the most comprehensive wave-optical treatment applicable for a wide class of realistic astrophysical lenses, especially those with an axisymmetric mass distribution.

\end{document}